\documentclass[final,a4paper,10pt]{article}

\usepackage{authblk}
\usepackage{hyperref}
\usepackage{subcaption}
\usepackage{amsmath}
\usepackage{amssymb}
\usepackage{graphicx}
\usepackage{fancyvrb}
\usepackage{comment}
\usepackage{listing}
\usepackage{listings}
\usepackage{float}
\usepackage{tikz}
\usepackage{array}
\usepackage{tabularx}
\usepackage{geometry}
\usepackage{xurl}
\usetikzlibrary{decorations.pathreplacing}
\usetikzlibrary{decorations.pathreplacing,shapes,shadows,arrows,positioning,graphs}

\DeclareMathOperator{\tr}{tr}

\usepackage{xcolor}
\definecolor{mygreen}{RGB}{28,172,0} 
\definecolor{mylilas}{RGB}{170,55,241}
\definecolor{myazure}{RGB}{0,175,225}
\definecolor{myorange}{RGB}{175,125,0}
\definecolor{myart}{RGB}{255,0,0}
\definecolor{review}{RGB}{255,0,0}
\definecolor{myrev}{RGB}{255,0,0}

\definecolor{myart}{RGB}{0,0,0}
\definecolor{review}{RGB}{0,0,0}
\definecolor{myrev}{RGB}{0,0,0}

\newcolumntype{L}[1]{>{\raggedright\let\newline\\\arraybackslash\hspace{0pt}}m{#1}}
\newcolumntype{C}[1]{>{\centering\let\newline\\\arraybackslash\hspace{0pt}}m{#1}}
\newcolumntype{R}[1]{>{\raggedleft\let\newline\\\arraybackslash\hspace{0pt}}m{#1}}

\newcommand{\R}{\mathbb R}
\newcommand{\N}{\mathbb N}

\def\IntO{\int\limits_{\Omega}}

\newcommand\dxb{\, \mathrm{d}\textbf{x}}

\newcommand{\nb}{n_b}

\newcommand{\nt}{|\mathcal{T}|}
\newcommand{\nn}{|\mathcal{N}|}
\newcommand{\ntp}{\|\mathcal{T}\|}
\newcommand{\nnf}{|\mathcal{M}|}

\renewcommand{\dim}{dim}

\newcommand{\Z}{I_{LG}}
\newcommand{\dn}{I_{DN}}
\newcommand{\dnn}{\dn(\n)}

\renewcommand{\d}{d}
\newcommand{\ii}{i}
\renewcommand{\j}{j}
\renewcommand{\k}{k}
\renewcommand{\l}{\ell}
\newcommand{\m}{m}
\newcommand{\n}{n}
\newcommand{\g}{r}

\newcommand{\iset}{\ii \in \{1, \hdots, \nb\}}
\newcommand{\inset}{\ii \in \{1, \hdots, \nn\}}
\newcommand{\jset}{\j \in \{1, \hdots, d\}}
\newcommand{\kset}{\k \in \{1, \hdots, \nt\}}
\newcommand{\lset}{\l \in \{1, \hdots, dim+1\}}
\newcommand{\mset}{\m \in \{1, \hdots, dim\}}
\newcommand{\nset}{\n \in \{1, \hdots, d \nn\}}
\newcommand{\imset}{\ii \in \{1, \hdots, \nnf\}}
\newcommand{\gset}{\g \in \{1, \hdots, \nnf\}}
\newcommand{\nfset}{\n \in \{1, \hdots, \d \nnf\}}
\newcommand{\kpiset}{\k \in \{1, \hdots, |\mathcal{T}^{\ii}|}
\newcommand{\kpnset}{\k \in \{1, \hdots, |\mathcal{T}^{(\n)}|}

\renewcommand{\b}{\textbf{b}}
\renewcommand{\u}{\textbf{u}}
\newcommand{\vv}{\textbf{v}}
\newcommand{\uu}{\textbf{u}}

\newcommand{\x}{\textbf{x}}
\newcommand{\F}{\textbf{F}}

\newcommand{\e}{\textbf{e}}
\newcommand{\I}{\textbf{I}}

\newcommand{\B}{\textbf{B}}

\newcommand{\M}{\textbf{M}}
\newcommand{\V}{\textbf{V}}

\newcommand{\T}{\mathcal{T}}
\newcommand{\f}{\textbf{f}}

\newcommand{\phib}{\boldsymbol{\varphi}}
\newcommand{\fjm}{\F_{\j,\m}}
\newcommand{\dfjmvn}{\frac{\partial \fjm}{\partial v_{\n}}}
\newcommand{\fjmsub}{\F_{\j,\m}^{sub}}
\newcommand{\Fv}{\F(\vv)}
\newcommand{\WFv}{W\big(\Fv\big)}

\newcommand{\vj}{v^{(\j)}}
\newcommand{\vjx}{v^{(\j)}(\x)}

\newcommand{\vij}{\vj_{\ii}}

\newcommand{\fx}{\f(\x)}

\newcommand{\vx}{\vv(\x)}

\newcommand{\fj}{\f^{(\j)}}
\newcommand{\en}{\e_{\n}}

\newcommand{\Jv}{J(\vv)}
\newcommand{\Ju}{J(\uu)}

\newcommand{\Jgrad}{J_{grad}}
\newcommand{\Jgradv}{\Jgrad(\vv)}
\newcommand{\Jgradu}{\Jgrad(\uu)}
\newcommand{\Jgradp}{\Jgrad^{(\n),+}(\vv)}
\newcommand{\Jgradm}{\Jgrad^{(\n),-}(\vv)}
\newcommand{\Jlin}{J_{lin}}
\newcommand{\Jlinv}{J_{lin}(\vv)}

\newcommand{\Tn}{\T^{(\n)}}
\newcommand{\Ti}{\T^{\ii}}

\newcommand{\Tkn}{T_{\k}^{(\n)}}

\newcommand{\pii}{p_{\ii}}

\newcommand{\textbfn}[1]{'\textbf{#1}'}

\newtheorem{remark}{Remark}
\newtheorem{example}{Example}
\newtheorem{benchmark}{Benchmark}

\urldef{\mailsb}\path|jan.valdman@utia.cas.cz |
\urldef{\mailam}\path|alexmos@ntis.zcu.cz |
\urldef{\mailcm}\path|matonoha@cs.cas.cz |

\makeatletter  
\renewcommand{\verbatim@font}{%
  \ttfamily\small\catcode`\<=\active\catcode`\>=\active
}
\makeatother   

\tikzset{
    circ/.style = {%
        draw, solid,
        fill = white,
        shape = circle,
        inner sep = 1pt,
        node contents =
    }
}

\title{Fast MATLAB evaluation of nonlinear energies \\
using FEM in 2D and 3D: Nodal elements}

\author[1]{Alexej Moskovka}
\author[2,3]{Jan Valdman}
\affil[1]{Department of Mathematics, Faculty of Applied Sciences,
University of West Bohemia, Technick\' a 8, 30100 Plze\v n, Czechia}
\affil[2]{Institute of Information Theory and Automation,
Czech Academy of Sciences, Pod vod\'{a}renskou
v\v{e}\v{z}\'{\i}~4, 18200~Praha~8, Czechia}
\affil[3]{Department of Computer Science, Faculty of Science,
University of South Bohemia, Brani\v{s}ovsk\'{a} 1760,
37005~\v{C}esk\'{e}~Bud\v{e}jovice, Czechia  }
\date{}                     

\begin{document}
\maketitle
\begin{abstract}
Nonlinear energy functionals appearing in the calculus of variations can be discretized by the finite element (FE) method and formulated as a sum of energy contributions from local elements. A fast evaluation of energy functionals containing the first order gradient terms is a central part of this contribution. We describe a vectorized implementation using the simplest {\color{review}linear nodal (P1)} elements in which all energy contributions are evaluated all at once without the loop over triangular or tetrahedral elements. Furthermore, in connection to the first-order optimization methods, the discrete gradient of energy functional is assembled in a way that {\color{myart}the} gradient components are evaluated over all degrees of freedom all at once. The key ingredient is the vectorization of exact or approximate energy gradients over nodal patches. It leads to a time-efficient implementation at higher memory-cost. Provided codes in MATLAB related to 2D/3D hyperelasticity and 2D p-Laplacian problem are available for download and structured in a way it can be easily extended to other types of vector or scalar forms of energies.
\end{abstract}

\section{Introduction}\label{sec:intro}
Given a domain $\Omega \in \R^{\dim}$, where $\dim \in \{1, 2, 3 \}$ is the space-dimension, we consider a minimization problem 
\begin{equation} \label{minimization_general}
J(\u)=\min_{\vv \in V} \Jv,
\end{equation}
where $V$ is a space of trial functions and $J: V \rightarrow \R$
represents {\color{myrev}an energy functional } in the form
\begin{equation} \label{hyperelasticity}
\Jv= \Jgradv + \Jlinv,
\end{equation}
where  $\Jgradv$ denotes its first-gradient part and $\Jlinv$ its linear part. Examples of such minimization problems are numerous and their study is a general subject of {\color{myart}the} Calculus of variations. There are models with higher order derivatives (such as plate problems with the second derivative in their formulation) available but not considered in this contribution. 

As the main example we recall a class of vector nonlinear elasticity problems represented by minimizations of energies of hyperelastic materials \cite{KrRo, MaHu}.  The trial space is chosen as 
$$ V = W^{1,p}_D(\Omega, \R^{\dim}) $$
i.e., the (vector) Sobolev space of $L^p(\Omega)$ integrable functions with the first weak derivative being also $L^p(\Omega)$ integrable and satisfying (in {\color{myart}the} sense of traces) Dirichlet boundary conditions $\vx=\u_D(\x)$ at the domain boundary $\x \in \partial \Omega$ for {\color{myart}a} prescribed function $\u_D: \partial \Omega \rightarrow \R^{\dim}$. A~ primal variable is {\color{myart}the} deformation mapping $\vv \in V$ describing the relocation of any point $\x \in \Omega$ during the deformation process. 
{\color{myrev}Then the gradient deformation tensor $\F \in L^p(\Omega,\R^{\dim \times \dim})$ is defined as}
\begin{equation} \label{deformation}
\F(\vv) = \nabla \vv = 
\begin{bmatrix}
\frac{\partial v^{(1)}}{\partial x_1} & \dots & \frac{\partial v^{(1)}}{\partial x_{\dim}} \\
\vdots && \vdots\\
\frac{\partial v^{(\dim)}}{\partial x_1} & \dots &\frac{\partial v^{(\dim)}}{\partial x_{\dim}} \\
\end{bmatrix}.
\end{equation}
{\color{myrev}The first-gradient and the linear parts of the energy functional \eqref{hyperelasticity} read }
\begin{equation*}
    \Jgradv = \IntO W\big(\F(\vx)\big) \dxb, \qquad \Jlinv= -\IntO  \fx \cdot \vx \dxb \, ,
\end{equation*} 
{\color{review}where $W : \R^{\dim \times \dim} \rightarrow \R$ defines a strain-energy density function  and $\f: \Omega \rightarrow  \R^{\dim} $ a loading functional. We assume the compressible Neo-Hookean density}
\begin{equation} \label{neoHook}
    W(\F) = C_1 \big(I_1(\F)-\dim -2  \log(\det \F)\big) + D_1 (\det \F -1)^2,
\end{equation}
where $I_1(\F)=|\F|^2$ uses the Frobenius norm $|\cdot |$, {\color{myrev} and
$\det(\cdot)$ is the matrix determinant operator}.  An extension to other gradient densities as the St. Venant Kirchhoff is possible. 
 
As the second example we recall {\color{myart}a} scalar p-Laplacian problem \cite{Drabek} with the energy functional defined as
\begin{equation}\label{pLaplacian}
J(v)=\frac{1}{p} \IntO |\nabla v(\x)|^p \dxb - \IntO  f(\x) \, v(\x)  \dxb \, ,
\end{equation}
where $V= W^{1,p}_D(\Omega, \R)$. The functional $J(v)$ is then known to be strictly convex in $V$ for {\color{review}$p \in (1,\infty)$} and it has therefore a unique minimizer $u(\x) \in V$. 

The main motivation of this contribution is to describe how nonlinear energy functionals can be efficiently and automatically evaluated by the finite element method (FE). We provide vectorization concepts and MATLAB implementation {\color{myrev}for 
\begin{itemize}
 \item evaluations of the value $\Jv$,
    \item evaluations of the gradient vector $\nabla \Jv$   
\end{itemize}
}
\noindent
expressed for a trial function $\vv \in V$. These two objects can be passed to an external optimization method of the first order to find a minima of the energy $\uu \in V$ and the corresponding minimal energy value $\Ju$. We utilize the trust-region optimization method \cite{conn2000} available in the MATLAB Optimization Toolbox for benchmarking. Our implementation is built on the top of {\color{review}our} own vectorized codes \cite{AnjamValdman2015,CSV2019,RahmanValdman2013} developed primarily for assemblies of finite element matrices. There are no explicit loops over mesh elements in evaluations of $\Jv$ and $\nabla \Jv$, all necessary data such as gradients of basis functions and energy densities are computed all at once. It leads to {\color{myart}a} significant computational speedup, but also it is memory intensive. Practically, the user specifies the form of the energy $\Jv$ and the corresponding gradient vector $\nabla \Jv$ is evaluated approximately {\color{myrev}by} a central difference scheme. Alternatively, if the user is willing to apply some differential calculus to a particular form of the energy, then the exact gradient can be assembled. This exact gradient approach is not versatile, but leads to {\color{myart}the} further performance speedup in our tests. {\color{myrev}We set up a set of six benchmarks as a base for future tests and improvements}:
\begin{itemize}
    \item {\color{myart}The} mesh and {\color{myart}the} nodal patches data {\color{myrev}are} preprocessed in Benchmark 1.  
    \item Evaluations of $\Jv$ and $\nabla \Jv$ for a given $\vv \in V$ are provided in Benchmarks 2 and 3. 
    \item {\color{myart}The} full minimizations of 2D/3D hyperelasticity and 2D p-Laplacian energies are shown in Benchmarks 4, 5 and 6.  
\end{itemize}
Authors are not aware of any similar MATLAB implementation. This is also our first attempt in this direction {\color{myrev}apart from our own contribution \cite{MMV} focusing on p-Laplacian energy minimization}. There is a growing number of MATLAB vectorized implementations {\color{myrev}of the second order linear partial differential equations} eg. \cite{JoKo, WeSo} or particular nonlinear partial differential equations \cite{CSV2019, GKR}.

The paper is structured in the following way: Section \ref{sec:notation} summarizes useful notation and Section \ref{sec:fem} basics of FEM. Section \ref{sec:meshpatch} includes implementation of two structures: mesh and (nodal) patches. Section \ref{sec:energy} is focused on implementation of energy evaluation and Section \ref{sec:gradient} on implementation of the gradient of energy. The final Section \ref{sec:minimization} reports on {\color{myart}the} solutions of particular minimization problems.

\section{Notation}\label{sec:notation}


\noindent
{\color{review}Index mappings in the construction of FEM:
\begin{description}
\item $\Z : \N \rightarrow \N$ - (local to global) mapping which for a local basis function on an element returns the index of the corresponding global basis function
\item $\dn : \N \rightarrow \N$ - (degree to node) mapping which for the $\n$-th degree of freedom returns the index $\ii$ of the corresponding node, see \eqref{idn}
\end{description}}
\noindent
Domain triangulations are described by {\color{myrev}the following} parameters:
\begin{description}
\item $\mathcal{T}, \mathcal{N}$ - a set of elements, a set of nodes
\item $\mathcal{M} \subset \mathcal{N}$ - a set of (at least partially) free nodes
\item $\mathcal{T}^{\ii}, \mathcal{T}^{(\n)}$ - the $\ii$-th nodal patch, the $\big(\dnn\big)$-th nodal patch
{\color{myrev}\item $|\mathcal{T}|, |\mathcal{N}|, |\mathcal{M}|, |\mathcal{T}^{\ii}|, |\mathcal{T}^{(\n)}|$ - the sizes of sets $\mathcal{T}, \mathcal{N}, \mathcal{M}, \mathcal{T}^{\ii}, \mathcal{T}^{(\n)}$}
\end{description}
\noindent
Both scalar and vector problems are treated together as a vector problem with $d$ components, where $d=1$ for scalar problems, $d=\dim$ for vector problems and $\dim$ is the space dimension. The following indices are frequently used: 
\begin{description}
\item  $\ii$ - {\color{myart}the} index of {\color{myart}a} node ($\iset$, also $\inset$ or $\imset$)
\item $\j$ - {\color{myart}the} index of {\color{myart}a} vector component ($\jset$)
\item $\k$ - {\color{myart}the} index of {\color{myart}an} element ($\kset$), also ($\kpiset$) or ($\kpnset$)
\item $\l$ - {\color{myart}the} index of {\color{myart}a} local basis function ($\lset$)
\item $\m$ - {\color{myart}the} index of {\color{myart}a} spatial component ($\mset$)
\item $\n$ - {\color{myart}the} index of {\color{myart}a} global degree of freedom ($\nset$) or {\color{myart}an} active degree of freedom ($\nfset$)
\item $\g$ - {\color{myart}the} index of {\color{myart}a} global patches matrix row ($\gset$)
\end{description}
\noindent
Nodal basis functions are used in several ways:
\begin{description}
\item $\varphi_{\ii}(\x), \phib_{\ii}(\x)$ - a scalar global nodal basis function, a vector global nodal basis function
\item $\phib_{\k,\l}(\x)$ - {\color{myart}the} $\ell-$th local basis function on the $k-$th element 
\end{description}
\noindent A trial vector function is addressed in several ways: 
\begin{description}
\item $\vx, \, \vjx$ - a trial vector function and its $\j$-th component
\item $\V$ - a matrix of {\color{myart}the} coefficients of $\vx$ in the nodal finite element basis 
\item $\vv_{\ii} \,,\, \vv^{(\j)} \,,\, \vij$ - the $\ii$-th row of $\V$, the $\j$-th column of $\V$, the (i,j) element of $\V$
\item $\vv$ - a vector reshaped from $\V$
\item $\hat{\vv}$ - the restriction of $\vv$ to free nodes
\end{description}
\noindent
Given matrices  $\textbf{A}, \textbf{B} \in \R^{p \times q},$ the following operators are used:
\begin{description}
\item $\tr(\textbf{A})$ - the trace of matrix {\color{myrev}(for $p = q$)}
\item $\det(\textbf{A})$ - the determinant of matrix  {\color{myrev}(for $p = q$)}
\item {\color{review}$\textbf{A} \odot \textbf{B}$ - the elementwise (Hadamard) product defined as a matrix $\textbf{C} \in \R^{p \times q}$, where $c_{i,j} = a_{i,j} \, b_{i,j}$}
\item {\color{review}$\textbf{A} : \textbf{B}$ - the scalar product defined as $\textbf{A} : \textbf{B} = \sum\limits_{i=1}^p \sum\limits_{j=1}^q a_{i,j} b_{i,j} = \sum\limits_{i=1}^p \sum\limits_{j=1}^q \textbf{A} \odot \textbf{B}$}
\end{description}

\section{Finite element discretization} \label{sec:fem}
The finite element method \cite{Ciarlet-FEM} is applied for {\color{myart}the} discretization of \eqref{minimization_general}.
We assume {\color{myart}a} trial function and {\color{myart}a} trial space of the form 
$$ \vx = (v^{(1)}(\x), \hdots, v^{(d)}(\x)), \qquad V = V^{(1)} \times \hdots \times V^{(d)}$$
\noindent
and approximate $\vx \in V$ in the finite-dimensional subspace 
$$V_h = V_h^{(1)} \times \hdots \times V_h^{(d)} \subset V, $$  
where $V_h^{(1)} = \hdots = V_h^{(d)} := V_h^s$
and {\color{myart}the} scalar basis space  $V_h^s$ is generated from {\color{myart}the} scalar basis functions 
$$\varphi_{\ii}(\x) \in V_h^s, \qquad \iset,$$ 
where $\nb$ denotes their number. Hence, any component of $\vx$ is given by a linear combination
\begin{equation} \label{linearcombsingle}
    \vj(\x) = \sum_{\ii=1}^{\nb} \vij \, \varphi_{\ii}(\x) \, , \qquad \x \in \Omega, \quad \jset .
\end{equation}
{\color{myart}The} coefficients $\vij$ from \eqref{linearcombsingle} are assembled in a matrix $\V \in \R^{\nb \times d}$ given as
\begin{equation} \label{vmatrix}
\V =
\begin{pmatrix}
v_1^{(1)} & \hdots & v_1^{(d)} \\
\vdots & & \vdots \\
\vdots & & \vdots \\
v_{\nb}^{(1)} & \hdots & v_{\nb}^{(d)}
\end{pmatrix}
=
\begin{pmatrix}
\vv_{1} \\
\vdots \\
\vdots \\
\vv_{\nb}
\end{pmatrix}
=
\begin{pmatrix}
\vv^{(1)} \dots \vv^{(d)} \\
\end{pmatrix}
\end{equation}
and latter two equivalent forms assume a row vector $\vv_{\ii} = (v_{\ii}^{(1)}, \hdots, v_{\ii}^{(d)}) \,,\, \iset$, and a column vector $\vv^{(\j)}=(v_1^{(\j)},\hdots, v_{\nb}^{(\j)})^{T},\, \jset$. Using a row vector basis function
$$ \phib_{\ii}(\x)  = \underbrace{(\varphi_{\ii}(\x), \hdots, \varphi_{\ii}(\x))}_{d \, -\mbox{ times}}, \qquad \iset  $$
one can rewrite \eqref{linearcombsingle} in a compact way for all components $\jset$ as
\begin{equation} \label{linearcombvector}
    \vx = \sum_{\ii=1}^{\nb}  \vv_{\ii} \odot \phib_{\ii}(\x) \, , \qquad x \in \Omega \, ,
\end{equation}
where the symbol $\odot$ represents an {\color{review}elementwise (or Hadamard) multiplication} (here multiplication of components with the same index $\j$). Formula \eqref{linearcombvector} is {\color{myart}the} key tool {\color{myrev}for} a vectorized implementation, since MATLAB provides the elementwise multiplication. 

The domain $\Omega$ is then approximated by its triangulation $\mathcal{T}$ into closed elements in the sense of Ciarlet \cite{Ciarlet-FEM}. The simplest possible elements are considered, i.e., 
triangles for $\dim=2$ and tetrahedra for $\dim=3$. The elements are geometrically specified by {\color{review}their} nodes (or vertices) belonging to the set of nodes $\mathcal{N}$. {\color{myart}The} nodes are also clustered into elements edges (for $\dim \geq 2$) and faces (for $\dim=3$). {\color{myart}The} numbers of elements and nodes are denoted as $\nt$ and $\nn$. 

Given a node $N_i \in \mathcal{N}, \inset$, we define a nodal patch {\color{myrev}$\Ti$ by 
$$\Ti = \{T  \in \mathcal{T}: N_{\ii} \in T \}$$} and the number of its elements by $|\Ti|$. The nodal patch $\Ti$ consists of elements denoted as $T_{\k}^{\ii} \,,\, \k \in \{1, \hdots, |\Ti|\}$, which are adjacent to the node $N_{\ii}$.
{\color{myrev}The same nodal patch $\Ti$ can be alternatively denoted by $\Tn$, where $\ii = \dn(\n)$ and $\n$ is the index of one of the corresponding degrees of freedom.
}

  We consider only the case where $V_h^s = P^1(\mathcal{T})$ is the space of nodal, elementwise linear and globally continuous scalar basis functions. Then {\color{myart}the} number of basis {\color{review}functions} is equal to {\color{myart}the} number of nodes
$$\nb =  \nn \, ,$$ 
but some coefficients of the trial function $\vx$ are known due to the Dirichlet boundary conditions.

\begin{example}
One regular triangulation $\T$ of {\color{myart}an} L-shape domain $\Omega$ is shown in Fig. \ref{example_2D}. The triangulation is specified by $\nt=24$ and $\nn=21$. The graph of the global scalar basis function $\varphi_{10}(\x)$ is displayed. The function has a hexagonal pyramid shape and {\color{myart}the} support on the nodal patch
$$\mathcal{T}^{10} = \{T_1, T_2, T_7, T_8, T_{19}, T_{20} \}.$$ {\color{myart}The} restrictions of $\varphi_{10}(\x)$ to its six supporting triangles in $\mathcal{T}^{10}$ are given as linear functions with values 1 at the node $N_{10}$ and values 0 at {\color{myart}the} two remaining nodes.
\begin{figure}[H]
\centering
\includegraphics[width=0.90\textwidth]{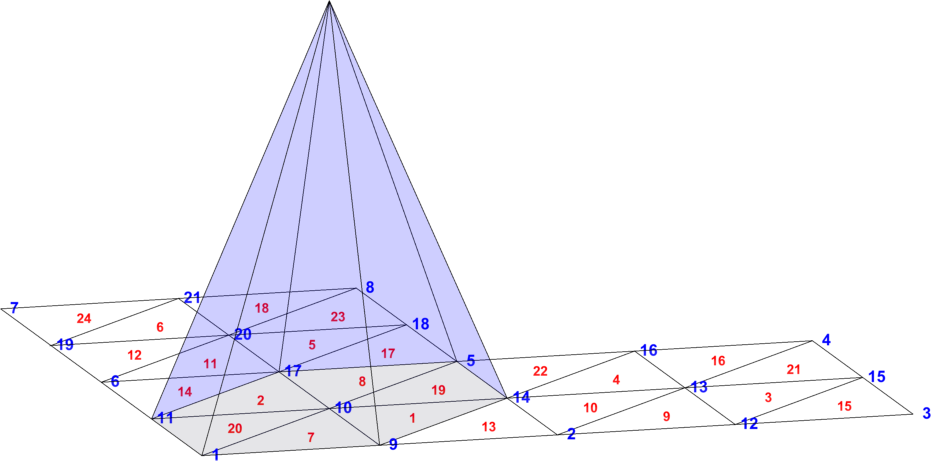}
\caption{Example of a scalar nodal basis function for $\dim=2$.  
}
\label{example_2D}
\end{figure}
\noindent
{\color{myrev}Additionally, for the scalar case ($\d = 1$) the node $N_{10}$ has only one corresponding degree of freedom with index $10$ and, therefore, $\T^{(10)} = \T^{10}$ as long as $\dn(10) = 10$. For the vector case ($d = 2$) the same node $N_{10}$ has two corresponding degrees of freedom with indices $19,20$, hence, $\T^{(19)} = \T^{(20)} = \T^{10}$ as long as $\dn(19) = \dn(20) = 10$. This alternative notation is essential in Section \ref{sec:gradient}.}
\end{example}

\subsection{The first-gradient energy term $\Jgradv$}
Since the first gradient of any scalar basis function $\varphi_{\ii}(\x) \in P^1(\mathcal{T}), \, \inset$, is a {\color{myrev}piecewise} constant function on each element, the gradient part of the discrete energy can be written as a sum over {\color{myart}the} elements
\begin{equation} \label{Denergy}
    \Jgradv = \IntO W\big(\F(\vx)\big) \dxb = \sum_{\k=1}^{\nt} \int_{T_{\k}} W\big(\nabla \vx\big) \dxb \, = \sum_{\k=1}^{\nt} |T_{\k}| \, W\big(\nabla \vx \big|_{{T_{\k}}}\big) \, ,
\end{equation}
and $|T_{\k}|$ denotes the size of the element $T_{\k}, \, \kset$ (equal to its length in 1D, its area in 2D or its volume in 3D). In order to evaluate \eqref{Denergy}, we need to assemble the gradient $\vx|_{T_{\k}}$ on every element. To do it, we define a (local-global) mapping $\Z : \N \times \N \rightarrow \N$ which for the $\k$-th element and its $\l$-th node returns the global index $\ii$ of this node. Then a local basis function is given as 
\begin{equation}
    \phib_{\k,\l}(x) = \phib_{\ii}\big|_{T_{\k}}(x), \qquad \mbox{where } i = \Z(\k,\l)
\end{equation}
for $\kset, \, \lset$. Hence, any partial derivative of $\vx$ with respect to $x_m$ reads
\begin{equation} \label{Dv}
        \frac{\partial \vx}{\partial x_{\m}} \bigg|_{T_{\k}} = \sum_{\l=1}^{\dim+1} \vv_{\k,\l} \odot \frac{\partial \phib_{\k,\l}(\x) }{\partial x_{\m}} \, , \qquad \mset \, ,
\end{equation}
where $\vv_{\k,\l}=\vv_{\Z(\k,\l)}$ represents {\color{myart}the} values of $\vv$ in the $\l$-th node of the $\k$-th element.

\subsection{The linear energy term $\Jlin$}

Furthermore, if $\fx \in V_h$, then the linear term of the energy \eqref{hyperelasticity} rewrites as
\begin{equation}
    \begin{split}
        \Jlinv = &\int_{\Omega} \fx \cdot \vx \dxb = \sum_{\j=1}^d \int_{\Omega} f^{(\j)}(\x) v^{(\j)}(\x) \dxb = \\ 
        &= \sum_{\j=1}^d \int_{\Omega} \bigg( \sum_{{\ii}_1=1}^{\nn} f_{\ii_1}^{(\j)} \, \varphi_{\ii_1}(\x) \sum_{\ii_2=1}^{\nn} v_{\ii_2}^{(\j)} \, \varphi_{\ii_2}(\x) \bigg) \dxb = \\
        &=\sum_{\j=1}^d \sum_{\ii_1=1}^{\nn} \sum_{\ii_2=1}^{\nn} f_{\ii_1}^{(\j)} v_{\ii_2}^{(\j)} \int_{\Omega} \Big( \varphi_{\ii_1}(\x)  \, \varphi_{\ii_2}(\x) \Big) \dxb \, .
    \end{split}
\end{equation}
All integral terms in the formula above can be assembled in a sparse and symmetric mass matrix  $\M \in \mathbb{R}^{\nn \times \nn}$ with entries
\begin{equation} \label{mass2D}
\M_{\ii_1,\ii_2}  = \int_{\Omega} \varphi_{\ii_1}(\x) \, \varphi_{\ii_2}(\x) \dxb,  \qquad \ii_1, \ii_2 \in \{1,2,\hdots, \nn\}.
\end{equation}
Then we can define vectors $\b^{(\j)} = \M \, \fj  \in \mathbb{R}^{\nn},$
where $\fj = (f_1^{(j)}, \hdots, f_{\nn}^{(j)})^T \in \mathbb{R}^{\nn},$ $ \jset $
and it is easy to check the exact formula
\begin{equation} \label{linearTerm2}
\Jlinv = \int_{\Omega}  \fx \cdot \vx \dxb = \sum_{\j=1}^{\d} \b^{(\j)} \cdot \vv^{(j)}
\end{equation}
which allows us to represent the linear part of the discrete energy $\Jv$ as a linear function.

\section{Implementation: Mesh and nodal patches} \label{sec:meshpatch}
{\color{myrev}
We describe the typical properties of meshes and nodal patches needed in our computational techniques. 
Considered finite element meshes consist of triangles (in 2D) and tetrahedra (in 3D), {\color{myrev}however} detailed explanations of Sections \ref{sec:meshpatch}, \ref{sec:energy}, \ref{sec:gradient} address 3D version only. }

\begin{figure}[H]
\centering
\begin{minipage}[t]{0.99\textwidth}
\includegraphics[width=\textwidth]{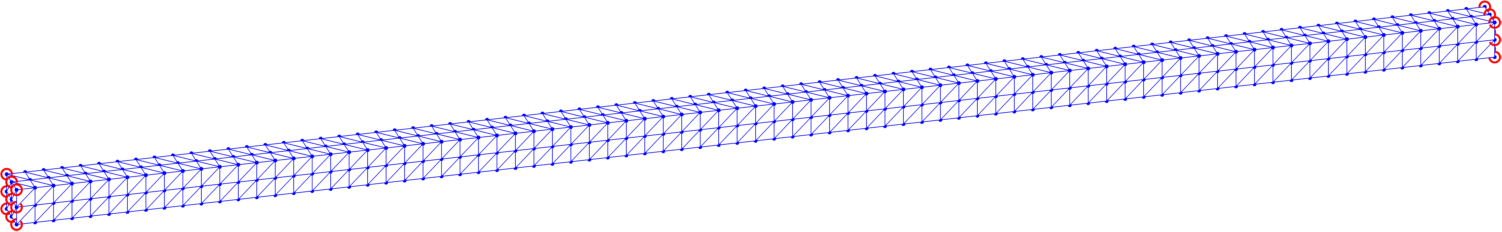}
\end{minipage}
\caption{Example: A tetrahedral mesh of a 3D bar domain with red boundary nodes.}
\label{mesh_torsion}
\end{figure}

\subsection{{\color{review}The \textbfn{mesh} structure}} \label{subsec:mesh}
The topology and important attributes of the computational domain are stored in {\color{myart}the} structure-type data object \textbfn{mesh}. For {\color{myart}the} example of {\color{myart}a} tetrahedral mesh displayed in Fig. \ref{mesh_torsion}, a vector energy model and full Dirichlet boundary conditions (specified in all three directions and indicated by {\color{myart}the} nodes in red circles) it provides the following information:

\begin{Verbatim}[xleftmargin=1cm]
mesh = 

  struct with fields:

                dim: 3
              level: 1
                 nn: 729
                 ne: 1920
        elems2nodes: [1920×4 double]
        nodes2coord: [729×3 double]
            volumes: [1920×1 double]
               dphi: {[1920×4 double] [1920×4 double] [1920×4 double]}
     nodesDirichlet: [18×1 double]
         nodesMinim: [711×1 double]
      dofsDirichlet: [54×1 double]
          dofsMinim: [2133×1 double]
\end{Verbatim}
\vspace{5mm}
Parameter \textbf{dim} represents the domain dimension (here $dim=3$) of the problem and \textbf{level} {\color{myart}the} level of {\color{myart}a} uniform refinement. Higher levels of refinement lead to finer uniformly refined meshes with more elements. {\color{myart}The} numbers of mesh nodes and elements are provided in \textbf{nn} and \textbf{ne}. Mesh nodes belonging to each tetrahedron are collected in a matrix \textbfn{elems2nodes} and {\color{myart}the} Cartesian coordinates of mesh nodes in a matrix \textbfn{nodes2coord}.

Once {\color{myart}the} last two matrices are given, the codes of \cite{RahmanValdman2013} generate {\color{myart}the} \textbfn{volumes} of all tetrahedra together with {\color{myart}the} restrictions of {\color{myart}the} partial derivatives (gradients) of all P1-basis functions to tetrahedra stored in a cell \textbfn{dphi} whose components are matrices corresponding to the partial derivates with respect to every $x_{\m} \,,\, \mset$.

{\color{myart}The} indices of Dirichlet boundary nodes are stored in a vector \textbfn{nodesDirichlet} and {\color{myart}the} remaining free nodes in a vector \textbfn{nodesMinim}. Vectors \textbfn{dofsDirichlet} and \textbfn{dofsMinim} {\color{review}denote} {\color{myart}the} indices of {\color{myart}the} degrees of freedom corresponding to Dirichlet and free nodes.

\begin{remark}
Vectors \textbfn{dofsDirichlet} and \textbfn{dofsMinim} are equal to vectors \textbfn{nodesDirichlet} and \textbfn{nodesMinim} {\color{myrev}in case of a scalar energy formulation.} 
\end{remark}

\subsection{{\color{review}The \textbfn{patches} structure}}
This object is only relevant if {\color{myart}the} knowledge of gradient $\nabla \Jv$ is required, e.g. as an input of the trust-region method of Section \ref{sec:minimization}. Then, {\color{myart}the} additional structure-type data object \textbfn{patches} is constructed for the evaluation of its gradient part given by \eqref{Denergy}.

\begin{remark} \label{remark:1}
Only the components of $\nabla \Jv$ corresponding to free nodes are evaluated in our implementation, and the remaining components belonging to the full Dirichlet boundary conditions are omitted. Thus, {\color{myart}the} nodes with at least one free degree of freedom also belong to the set of free {\color{myrev}nodes and the gradient} is evaluated in all {\color{myrev}of} their components and finally restricted to free degrees of freedom. 
\end{remark}

We denote by $\mathcal{M}$ {\color{myart}a} set of all free nodes and by $\nnf$ their number. Then, the node index  $\ii, \imset$, goes exclusively through the free nodes. A nodal patch $\Ti \,,\, \imset$ is implemented as
a vector of elements indices '\textbf{elems\_}$\boldsymbol{\ii}$', a vector of their volumes '\textbf{volumes\_}$\boldsymbol{\ii}$', a matrix of {\color{myart}the} corresponding elements nodes stored as '\textbf{elems2nodes\_}$\boldsymbol{\ii}$', values of {\color{myart}the} gradients of local basis functions stores as a cell '\textbf{dphi\_}$\boldsymbol{\ii}$' with matrices components {\color{review}\textbf{dphi\_}$\boldsymbol{\ii}\{1\}$, $\hdots$, \textbf{dphi\_}$\boldsymbol{\ii}\{\dim\}$}. All these matrices and vectors are of size $|\Ti| \,\times\, (\dim+1)$ and $|\Ti| \,\times\, 1$, respectively. 

\begin{figure}[h]
\centering
\begin{minipage}{0.5\textwidth}
\centering
\begin{tikzpicture}
\filldraw[thick, top color=white,bottom color=gray!30!] (0,0) rectangle node{$\mathcal{T}^{|M|}$} +(2, 3);

\draw (1,4.0) node[below] {$\vdots$} ;

\filldraw[thick, top color=white,bottom color=gray!30!] (0,4) rectangle node{$\mathcal{T}^{1}$} (2,7);
\draw [decorate,decoration={brace,amplitude=10pt},xshift=-4pt,yshift=0pt]
(0,0) -- (0,7.0) node [black,midway,xshift=-1.3cm] {{\footnotesize $\ntp$} rows};

\draw [decorate,decoration={brace,amplitude=10pt,mirror,raise=4pt},yshift=0pt]
(2,0) -- (2,3) node [black,midway,xshift=1.6cm] {{\footnotesize $|\mathcal{T}^{|M|}|$} rows};

\draw [decorate,decoration={brace,amplitude=10pt,mirror,raise=4pt},yshift=0pt]
(2,4) -- (2,7) node [black,midway,xshift=1.5cm] {{\footnotesize $|\mathcal{T}^{1}|$} rows};

\draw [stealth-](2.5,0.1) -- (3,0.1) node [black,midway,xshift=0.8cm]{$p_{|M|}$};
\draw [stealth-](2.5,2.9) -- (3,2.9) node [black,midway,xshift=1.3cm]{$p_{|M|-1}+1$};
\draw [stealth-](2.5,4.1) -- (3,4.1) node [black,midway,xshift=0.6cm]{$p_{1}$};
\draw [stealth-](2.5,6.9) -- (3,6.9) node [black,midway,xshift=0.5cm]{$1$};

\end{tikzpicture}

\end{minipage}
\caption{All data from nodal patches 
are stored in long matrices or vectors.} \label{pic:patches}
\end{figure}
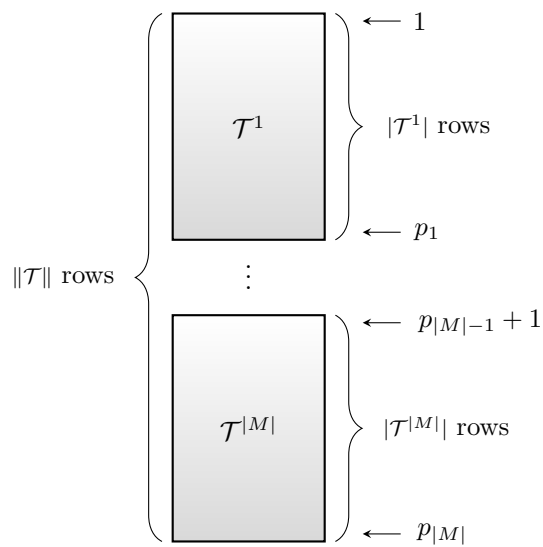

The data of all nodal patches $\mathcal{T}^{\ii} \,,\, \imset$ are then collected in {\color{myart}the} {\color{myrev}corresponding} long global matrices or vectors
with the number of rows equal to 
$$\ntp = \sum\limits_{\ii=1}^{\nnf} |\Ti| \, .$$
{\color{myrev}For $\imset$ we define {\color{myart}the} indices}
\begin{equation} \label{pind1}
    \pii = \sum\limits_{\g=1}^{\ii} |\mathcal{T}^{\g}| ,
\end{equation}
and additionally $p_0 = 0$. Then the submatrix or subvector extracted from rows $(p_{\ii-1}+1),\hdots,\pii$ of the global matrices or vector above corresponds to the $\ii$-th nodal patch. 
It is shown schematically in Fig. \ref{pic:patches}.

\noindent
Thus we obtain the global vectors \textbfn{elems}, \textbfn{volumes} of size $\ntp \,\times\, 1$ and the global matrices \textbfn{elems2nodes}, 
{\color{review}\textbf{dphi}$\{1\}$, $\hdots$, \textbf{dphi}$\{\dim\}$}, of size $\ntp \,\times\, (\dim+1)$.

\begin{remark}
For the gradient evaluation of Section \ref{sec:gradient} we will need to extend {\color{myart}a} set of $p_{\ii} \,,\, \ii \in \{0,\hdots,\nnf\}$, indices up to $\{0,\hdots,\d \nnf\}$. {\color{myrev}Thus, we additionally define
\begin{equation} \label{pind2}
    \begin{split}
        p_{\n} &= p_{\n\,-\,\nnf} \, + \;\, \ntp \, , \qquad \n \in \{\;\,\nnf+1,\hdots,2\nnf\} \\
        p_{\n} &= p_{\n-2\nnf} + 2\ntp \, , \qquad \n \in \{2\nnf+1,\hdots,3\nnf\} \, .
    \end{split}
\end{equation}
}
\end{remark}
 
In order to maintain the right ordering of {\color{myart}the} local basis functions within each nodal patch, {\color{myart}an} additional logical-type matrix of zeros and ones \textbfn{logical} is provided. If the $\n$-th row corresponds to the $\ii$-th patch, then \textbf{logical}$(\n,\l) = 1$ means that \textbf{elems2nodes}$(\n,\l) = \ii$. Therefore, in every row of \textbfn{logical} matrix {\color{myart}the} value '$1$' has exactly one single occurrence.

Below we provide an example of {\color{myart}the} \textbfn{patches} structure along with the same \textbfn{mesh} one introduced in Subsection \ref{subsec:mesh} and corresponding to the domain in Fig. \ref{mesh_torsion}.

\begin{Verbatim}[xleftmargin=1cm]
patches = 

  struct with fields:

        lengths: [711×1 double]
          elems: [7584×1 double]
        volumes: [7584×1 double]
    elems2nodes: [7584×4 double]
           dphi: {[7584×4 double]  [7584×4 double]  [7584×4 double]}
        logical: [7584×4 logical]
\end{Verbatim}
\vspace{5mm}
The first vector \textbfn{lengths} is of size $\nnf \times 1$ with entries $\textbf{lengths}(\ii) = |\Ti| \,,\, \imset$. Here, its size is equal to $711 = 729 - 18$, where $\nn = 729$ is the number of mesh nodes and $18$ is the number of nodes with full Dirichlet boundary conditions.

\begin{benchmark}\label{benchmark1}
The script {\color{myrev}\verb!benchmark1_start.m!} generates a sequence of {\color{myart}the} structures \verb!mesh! and \verb!patches! corresponding to each of {\color{myart}the} uniform mesh refinements. Table \ref{tab1} provides {\color{myart}the} assembly times and {\color{myart}the} memory requirements of both objects. 

\newcolumntype{d}{>{\hsize=1\hsize}X}
\newcolumntype{Y}{>{\raggedleft\arraybackslash}X}
\newcolumntype{s}{>{\hsize=.5\hsize}Y}
\newcolumntype{z}{>{\hsize=1.1\hsize}Y}
\newcolumntype{Z}{>{\centering\arraybackslash}X }
\begin{table}[H]
    \centering
      \begin{tabularx}{0.95\textwidth}{ s |Y |z |Y |Y |Y |z |z }
   mesh level & number of nodes $\nn$ & number of elems. $\nt$ & number of free dofs & \textbf{mesh} setup time [s] & \textbf{patches} setup time [s] & \textbf{mesh} memory size [MB] & \textbf{patches} memory size [MB]\\
     \hline 
 1 & 729 & 1920 & 2133 &      0.01 &      0.01 &      0.30 &      1.13 \\ 
 2 & 4025 & 15360 & 11925 &      0.02 &      0.02 &      2.32 &      9.07 \\ 
 3 & 26001 & 122880 & 77517 &      0.06 &      0.13 &     18.17 &     72.73 \\ 
 4 & 185249 & 983040 & 554013 &      0.51 &      0.98 &    144.07 &    582.53 \\ 
 5 & 1395009 & 7864320 & 4178493 &      5.56 &     10.62 &   1147.67 &   4663.18 \\ 
    \end{tabularx}
    \vspace{0.1cm}
    \caption{Benchmark 1 - setup times and memory consumption in 3D.}\label{tab1}
\end{table}
\noindent
Note that the most memory consuming part of both structures
is given by substructures \textbfn{dphi} containing {\color{myart}the} values of {\color{myart}the} precomputed gradients of basis functions.
\end{benchmark}



\section{Implementation: energy evaluation} \label{sec:energy}
The following matrix-vector transformation is frequently used: matrices $\B, \V \in \R^{\nn \times d}$ are stretched to {\color{myart}the} isomorphic vectors 
\begin{equation} \label{bvvectors}
    \begin{split}
       \b, \vv \in \R^{d \nn}: \qquad  \b_{\n} = \B_{\ii,\j}, \quad  \vv_{\n} = \V_{\ii,\j}, \qquad  \nset, 
    \end{split}
\end{equation}
where $\ii = (\n-1)/d \, + 1$ and $\j = (\n-1)\%d \, + 1$. Here, $/$ symbol is the integer division operator and $\%$ is the modulo operator. Put simply, for any $\inset$ the elements of $\vv$ with indices $d(\ii-1) +1, \hdots, d(\ii-1) +d$ corresponds to the values of the trial function $\vx$ in the $\ii$-th node in {\color{myart}the} directions $1, \hdots, d$. 

\subsection{The linear energy term $\Jlin$}
The linear part of the energy \eqref{linearTerm2} rewrites  equivalently as
\begin{equation} \label{linearTerm3}
    \Jlinv = \int_{\Omega}  \fx \cdot \vx \dxb = \B : \V = \b \cdot \vv, 
\end{equation}
where $:$ denotes the scalar product of matrices. 

\subsection{The first-gradient energy term $\Jgrad$} \label{subs:fgenergy}
The gradient part of the discrete energy \eqref{Denergy} is given as {\color{myart}a} sum of {\color{myart}the} energy contributions from every element $T_{\k}, \, \kset$ and its evaluation in MATLAB is performed effectively by using operations with vectors and matrices only.
The energy evaluation for a trial vector $\vv \in \R^{\d \nn}$ is performed by the main function:

\begin{listing}
\begin{lstlisting}
function [e, densities] = energy(v,mesh,params)
% components of deformation
v_cell = createCellFromVector(v,mesh.dim);   

% components of deformation on elements
v_elems = CellAtMatrixOfIndices(v_cell,mesh.elems2nodes); 

% deformation gradients on elements
F_elems = evaluate_F(mesh,v_elems);

% gradient densities on elements
densities.Gradient = densityGradientVector_3D(F_elems,params);

% total gradient energy
e = sum(mesh.volumes.*densities.Gradient);
end
\end{lstlisting}
\end{listing}

\noindent
The structure \textbfn{mesh} is described in Section \ref{sec:energy} and {\color{myart}the} \textbfn{params} contains material parameters apart from some other parameters (e.g. visualization parameters). The code above is vectorized and generates the objects:
\begin{description}
\item A cell \textbfn{v\_elems} containing matrices \textbf{v\_elems$\{1\}$}, \textbf{v\_elems$\{2\}$}, \textbf{v\_elems$\{3\}$} of size $\nt \, \times \, 4$
providing {\color{myart}the} restrictions of nodal deformations to all elements.
    \item A cell '$\F$\textbf{\_elems}' of size $\dim \times \dim$ storing the deformation gradients {\color{review}(see \eqref{deformation})} in all elements. In particular, $\F$\textbf{\_elems}$\{\d\}\{\m\}$ is then a vector of size $\nt \times 1$
    evaluating {\color{myart}the} partial derivatives of the $\d$-th component of deformation with respect to the $\m$-th variable in all elements.
    \item A vector \textbfn{densities.Gradient} of size $\nt \times 1$ containing gradient densities in all elements.
    \item The energy \textbf{e} is given as {\color{myart}a} sum of {\color{myart}the} gradient energy contributions over {\color{myart}the} elements (the gradient part $\Jgradv$ given by \eqref{Denergy}) subtracted by the linear energy term $\Jlinv$ (given by \eqref{linearTerm2}).
\end{description}
The {\color{myrev}Neo-Hookean} density function $W=W(F)$ from \eqref{neoHook} is implemented as

\begin{listing}
\begin{lstlisting}
function densities=densityGradientVector_3D(F,params)
% determinant term
DET = F{1,1}.*F{2,2}.*F{3,3} + F{1,3}.*F{2,1}.*F{3,2} + ...
      F{1,2}.*F{2,3}.*F{3,1} - F{1,3}.*F{2,2}.*F{3,1} - ...
      F{1,2}.*F{2,1}.*F{3,3} - F{1,1}.*F{2,3}.*F{3,2};

% I1 term (Frobenius norm squared)
I1 = F{1,1}.^2 + F{1,2}.^2 + F{1,3}.^2 + ...
     F{2,1}.^2 + F{2,2}.^2 + F{2,3}.^2 + ...
     F{3,1}.^2 + F{3,2}.^2 + F{3,3}.^2;

% gradient densities
densities = params.C1*(I1-3-2*log(DET)) + params.D1*(DET-1).^2;
end
\end{lstlisting}
\end{listing}

\begin{benchmark}\label{benchmark2}
 Assume {\color{myart}a} bar domain $\Omega = (0,l_x) \times (-\frac{l_y}{2} \,,\, \frac{l_y}{2}) \times (-\frac{l_z}{2} \,,\, \frac{l_z}{2})$, where $l_x = 0.4, l_y = l_z = 0.01$, specified by the material parameters  $E = 2\cdot10^8$ (Young's modulus) and  $\nu = 0.3$ (Poisson's ratio) and deformed by the prescribed deformation $\vx=\vv(x,y,z)$ given by
\begin{equation} \label{displacement}
    \begin{split}
        v^{(1)}(x,y,z) &= \; \; \;x \, , \\
        v^{(2)}(x,y,z) &= \; \; \; \cos(\alpha \frac{x}{l_x}) \, y + \sin(\alpha \frac{x}{l_x}) \, z \, , \\
        v^{(3)}(x,y,z) &= -\sin(\alpha \frac{x}{l_x}) \, y + \cos(\alpha \frac{x}{l_x}) \, z  \, ,
    \end{split}
\end{equation}
or, equivalently, using matrix operations
\begin{equation}
    \begin{pmatrix}
        v^{(1)}(x,y,z) \\
        v^{(2)}(x,y,z) \\
        v^{(3)}(x,y,z)
    \end{pmatrix}
    =
    \begin{pmatrix}
        1 & 0 & 0 \\
        0 & \cos(\alpha \frac{x}{l_x}) & \sin(\alpha \frac{x}{l_x}) \\
        0 & -\sin(\alpha \frac{x}{l_x}) & \cos(\alpha \frac{x}{l_x})
    \end{pmatrix}
    \begin{pmatrix}
        x \\ y \\ z
    \end{pmatrix} \, .
\end{equation}
Here, $\alpha = 2\pi$ means that the right Dirichlet wall is twisted once around the x-axes (Fig. \ref{twist}). 
\begin{figure}[b]
\centering
\includegraphics[width=\textwidth]{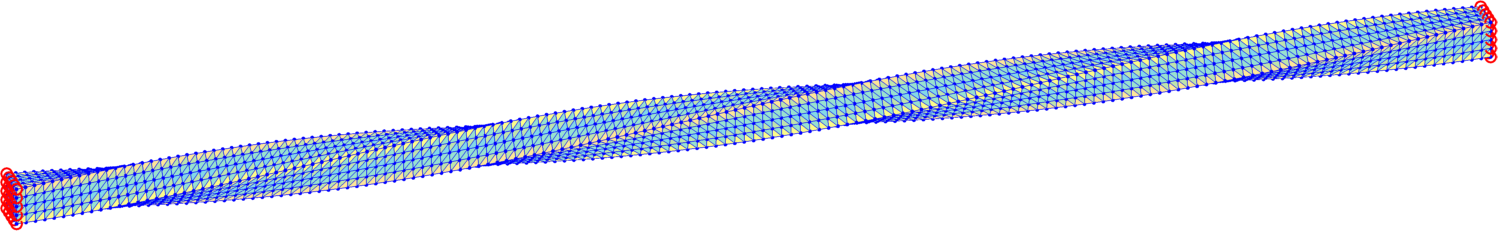}
\caption{A bar domain twisted by the prescribed deformation \eqref{displacement}.}
\label{twist}
\end{figure}

\noindent
Constants $C_1, D_1$ are transformed according to {\color{myart}the} formulas $C_1 = \frac{\mu}{2}, D_1 = \frac{K}{2},$ {\color{myrev}where $\mu = \frac{E}{2(1+\nu)}$ is the shear modulus and $K = \frac{E}{3(1-2\nu)}$ is the bulk modulus}. The exact evaluation shows that the corresponding gradient energy using the Neo-Hookean density function \eqref{neoHook} reads
{\color{review}$$ J_{grad}(\vv) = \frac{\alpha^2 \, C_1 \, l_y \, l_z \, (l_y^2 + l_z^2)}{12 \, l_x} \approx 6.326670.$$}
\noindent
The script {\color{myrev}\verb!benchmark2_start.m!} evaluates {\color{myart}an} approximation of $J_{grad}(\vv)$ on {\color{myart}the} sequence of {\color{myart}the} uniform mesh refinements defined in Benchmark \ref{benchmark1}. In order to provide {\color{myrev}higher accuracy of {\color{myart}the} evaluation times}, the energies are recomputed 10 times. Table \ref{tab2} provides evaluation times and values of the energy approximations {\color{review}(times for setting up the '\textbf{mesh}' structure are not included).}
\newcolumntype{d}{>{\hsize=1\hsize}X}
\newcolumntype{Y}{>{\raggedleft\arraybackslash}X}
\newcolumntype{s}{>{\hsize=.5\hsize}Y}
\newcolumntype{z}{>{\hsize=1.1\hsize}Y}
\newcolumntype{Z}{>{\centering\arraybackslash}X }
\begin{table}[h]
    \centering
    \begin{tabularx}{0.9\textwidth}
    {s |z |Y |Y }
   mesh level &  number of free dofs & evaluation (10x) of $\Jgradv$: time [s]  & value of $\Jgradv$  \\
     \hline 
 1 & 2133 &      0.01 &   12.6623 \\ 
 2 & 11925 &      0.02 &    7.9083 \\ 
 3 & 77517 &      0.08 &    6.7219 \\ 
 4 & 554013 &      0.76 &    6.4255 \\ 
 5 & 4178493 &     12.82 &    6.3514 \\ 
    \end{tabularx}
    \vspace{0.3cm}
    \caption{Benchmark 2 - energy evaluation times in 3D.}\label{tab2}
\end{table}
\end{benchmark}

\section{Implementation: energy gradient evaluation} \label{sec:gradient}

Evaluation of the full gradient $\nabla \Jv,$ where $\vv=(v_1, \dots, v_{\d \nn}) \in \R^{\d \nn}$, requires in general computation of all partial derivatives 
$\frac{\partial \Jv}{\partial v_n}$, $\nset$. {\color{myrev}Some} minimization methods (e.g. trust-region mentioned applied in Section \ref{sec:minimization}) require the knowledge of gradient, but only restricted to free degrees of freedom. 

The gradient $\nabla \Jlinv$
easily reads using \eqref{linearTerm3} 
\begin{equation} \label{linearTermGrad}
    \nabla \Jlinv = \b \, ,
\end{equation}
where $\b$ is of size $\d \nn$, given by \eqref{bvvectors} and its restriction to free degrees of freedom is then trivial.

The evaluation of $\nabla\Jgradv$ is {\color{review}technically more involved}. 
{\color{review}Firstly, it is evaluated with respect to Remark \ref{remark:1}
for all degrees of freedom belonging to all (at least partially) free nodes.
Secondly, it is restricted to free degrees of freedom only.} In order to determine to which node the $\n$-th active degree of freedom belongs we define the index mapping
\begin{equation} \label{idn}
\dn : \{1, \hdots, \d \nnf\} \rightarrow \{1, \hdots, \nnf\} , \qquad \dn(n) = (n-1)/d + 1
\end{equation}
which for the $\n$-th active degree of freedom returns the corresponding $i$-th free node (here $/$ is an integer division operator). {\color{myrev}By
$$\Tn \, , \qquad \nfset \, ,$$
we denote the set of elements adjacent to node $N_{\dnn}$ belonging to the $\n$-th active degree of freedom and by $|\Tn|$ their number.}
The gradient of the nonlinear part $\Jgradv$ can be computed in two different ways:
\begin{enumerate}
    \item numerically, where the partial derivatives are computed approximately by using a difference scheme.
    \item exactly by taking the explicit partial derivatives. 
\end{enumerate}
Deriving the exact partial derivatives can be demanding and it depends on the particular problem. On the contrary, the numerical approach is more general and is feasible regardless of the complexity of the function representing the corresponding discrete energy. Hence, we first describe the numerical approach by using the central difference scheme and then explain the gradient evaluation by deriving the explicit form of {\color{myart}the} partial derivatives.

\subsection{Numerical approach to evaluate $\nabla \Jgradv$} \label{subs:numeric}
By using {\color{myrev}the central difference scheme}, one can write
\begin{equation} \label{CD}
    \frac{\partial}{\partial v_{\n}} \Jgradv \approx \frac{\Jgrad(\vv + \varepsilon \en) - \Jgrad(\vv - \varepsilon \en)}{2 \varepsilon} \, ,
\end{equation}
where $\en$ is the $n$-th canonical vector in $\R^{\d \nnf}$ and 
$\varepsilon$ is a small positive number.
Both summands in the numerator above can be directly evaluated by taking the energy evaluation procedure introduced in the previous subsection as 
\begin{equation*} 
    \begin{split}
        &\Jgrad(\vv + \varepsilon \en) - \Jgrad(\vv - \varepsilon \en) = \\
        &= \sum\limits_{\k=1}^{\nt} \int_{T_{\k}} W(\nabla(\vv + \varepsilon \en)) \dxb -
          \sum\limits_{\k=1}^{\nt} \int_{T_{\k}} W(\nabla(\vv - \varepsilon \en)) \dxb \, . 
    \end{split}
\end{equation*}
However, this approach is ineffective as long as the step of {\color{myart}the} central difference scheme $\varepsilon$ occurs only in a few summands of the sums representing $\Jgrad(\vv + \varepsilon \en)$ and $\Jgrad(\vv - \varepsilon \en)$ given by \eqref{Denergy}, while the remaining summands are the same and therefore vanish. Hence, we can further simplify
\begin{equation} \label{CDS}
    \begin{split}
        &\Jgrad(\vv + \varepsilon \en) - \Jgrad(\vv - \varepsilon \en) = \\
        &= \sum\limits_{\k=1}^{|\Tn|} \int_{\Tkn} W(\nabla(\vv + \varepsilon \en)) \dxb - \sum\limits_{\k=1}^{|\Tn|} \int_{\Tkn} W(\nabla(\vv- \varepsilon \en)) \dxb = \\
        &= \sum\limits_{\k=1}^{|\Tn|} |\Tkn| W\big(\nabla(\vv + \varepsilon \en)\big|_{\Tkn}\big) - \sum\limits_{\k=1}^{|\Tn|} |\Tkn| W\big(\nabla(\vv - \varepsilon \en)\big|_{\Tkn}\big) \, .
    \end{split}
\end{equation}
By using the {\color{myrev}substitutions}
\begin{equation} \label{jgradpm}
    \Jgradm = |\Tkn| W\big(\nabla(\vv - \varepsilon \en)\big|_{\Tkn}\big) \, , \quad
    \Jgradp = |\Tkn| W\big(\nabla(\vv + \varepsilon \en)\big|_{\Tkn}\big) \, ,
\end{equation}
one can rewrite \eqref{CDS}
\begin{equation} \label{sumpatch}
    \Jgrad(\vv + \varepsilon \en) - \Jgrad(\vv - \varepsilon \en) = \sum\limits_{\k=1}^{|\Tn|} \Jgradp - \sum\limits_{\k=1}^{|\Tn|} \Jgradm
\end{equation}
  and evaluate the whole $\nabla \Jgradv$ via the simple for-loop over its components. 

\begin{remark}
The energy evaluation procedure has to be called in every loop over {\color{myart}the} vector components $\nfset$ and it turned out to cause multiple self built-in times that slowed down performance. To avoid that, the original energy evaluation procedure is modified so that {\color{myart}the} multiple input vectors can be processed simultaneously and the energy evaluation procedure be called only once.
\end{remark}
The outer gradient evaluation procedure is simple:

\begin{listing}
\begin{lstlisting}
eps = 1e-8;   % finite difference step size

% local patch energies for +eps and -eps
es_minsplus = energies(v,[-eps eps],mesh,patches,params,indx);

% the gradient vector by the central difference scheme
g = diff(es_minsplus,1,2)/eps/2;
\end{lstlisting}
\end{listing}
Here \textbfn{es\_minsplus} is a matrix of size $3\ntp \, \times \, 2$ with {\color{myart}the} components given by \eqref{jgradpm}:
\begin{center}
\textbf{es\_minsplus}$(\n,1) = \Jgradm$ , \qquad
\textbf{es\_minsplus}$(\n,2) = \Jgradp$ . \\
\end{center}
 The gradient vector \textbf{g} is then assembled by using the central difference scheme \eqref{CD} with a constant central difference step size $\varepsilon$ for all $\nfset$. Obviously, a generalization of the above code to higher accuracy difference {\color{review}schemes is} possible.

We recall a cell \textbfn{v\_elems} containing matrices
\begin{center}
\textbf{v\_elems$\{1\}$}, \textbf{v\_elems$\{2\}$}, \textbf{v\_elems$\{3\}$} \quad of size $\nt \times 4$
\end{center}
is assembled
in the energy evaluation procedure \textbf{energy} from Subsection \ref{subs:fgenergy}.
For {\color{myart}the} {\color{review}evaluation of \eqref{CD}} by using \eqref{sumpatch} we need to assemble a cell-structure \textbfn{v\_patches} containing matrices
\begin{center}
\textbf{v\_patches$\{1\}$}, \textbf{v\_patches$\{2\}$}, \textbf{v\_patches$\{3\}$} \quad of size $\ntp \, \times \, 4$
\end{center}
that provide {\color{myart}the} restrictions of nodal deformations to all patches. {\color{review} In fact, the structure \textbfn{v\_patches} copies parts of \textbfn{v\_elems} to {\color{myart}the} particular positions.}

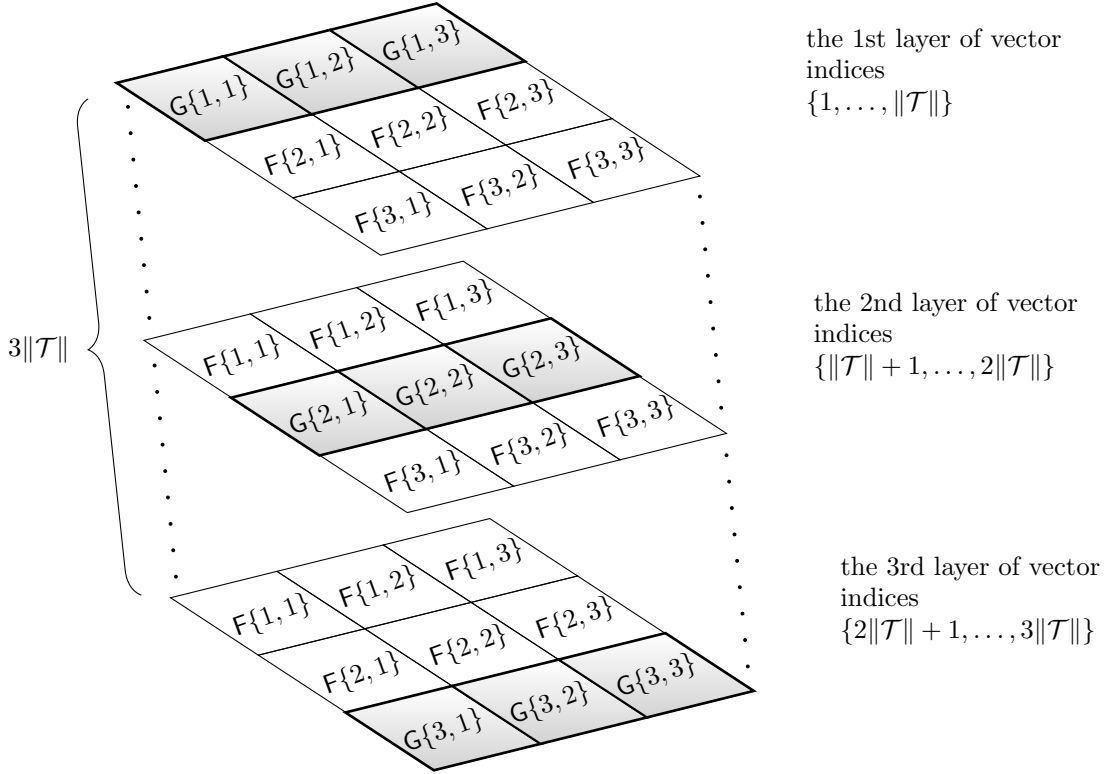
\begin{figure}[H]
\begin{tikzpicture}[scale=.7,every node/.style={minimum size=1cm},on grid]
    
    \def\a{2} \def\b{1.5}
    \def\xstart{0} \def\ystart{0}
    \def\X{5} \def\Y{5}
    
    \def\xtexts{0.05} \def\ytexts{3.85}
    \def\at{2} \def\bt{0.52}
    \def\aat{1.7} \def\bbt{1.1}
    \def\fx{0.45} \def\fy{0.2}
    \def\gx{0.45} \def\gy{0.2}
    \def\Xt{0.55} \def\Yt{4.9}
    \def\xnew{0.1}
    \def\rot{20}
    
    \def\r{0.75pt}
    \def\xdota{-4.75} \def\ydota{5.95}
    \def\xdotb{6} \def\ydotb{4.25}
    \def\xdotc{0.1} \def\ydotc{2.8}
    \def\dw{0.05} \def\dh{0.5}
    \def\Xc{0.5} \def\Yc{4.9}
    
    \def\xmain{2} \def\ymain{14}
    \def\xlabels{8.5} \def\ylabels{6.5}
    \def\al{0.5} \def\bl{5}
    
    \def\bg{35}
    
    \begin{scope}[
    	yshift=90,every node/.append style={
    	    yslant=0.25,xslant=-1.1},yslant=0.25,xslant=-1.1
    	  ]
        
        \draw[black] (\xstart+\X,\ystart+\Y) rectangle (\xstart+\a+\X,\ystart+\b+\Y);
        \draw[black] (\xstart+\a+\X,\ystart+\Y) rectangle (\xstart+2*\a+\X,\ystart+\b+\Y);
        \draw[black] (\xstart+2*\a+\X,\ystart+\Y) rectangle (\xstart+3*\a+\X,\ystart+\b+\Y);
        \draw[black] (\xstart+\X,\ystart+\b+\Y) rectangle (\xstart+\a+\X,\ystart+2*\b+\Y);
        \draw[black] (\xstart+\a+\X,\ystart+\b+\Y) rectangle (\xstart+2*\a+\X,\ystart+2*\b+\Y);
        \draw[black] (\xstart+2*\a+\X,\ystart+\b+\Y) rectangle (\xstart+3*\a+\X,\ystart+2*\b+\Y);
        \draw[black,very thick] (\xstart+\X,\ystart+2*\b+\Y) rectangle (\xstart+\a+\X,\ystart+3*\b+\Y);
        \draw[black,very thick] (\xstart+\a+\X,\ystart+2*\b+\Y) rectangle (\xstart+2*\a+\X,\ystart+3*\b+\Y);
        \draw[black,very thick] (\xstart+2*\a+\X,\ystart+2*\b+\Y) rectangle (\xstart+3*\a+\X,\ystart+3*\b+\Y);
        \filldraw[thick, top color=white,bottom color=gray!\bg!] (\xstart+\X,\ystart+2*\b+\Y) rectangle (\xstart+\a+\X,\ystart+3*\b+\Y);
        \filldraw[thick, top color=white,bottom color=gray!\bg!] (\xstart+\a+\X,\ystart+2*\b+\Y) rectangle (\xstart+2*\a+\X,\ystart+3*\b+\Y);
        \filldraw[thick, top color=white,bottom color=gray!\bg!] (\xstart+2*\a+\X,\ystart+2*\b+\Y) rectangle (\xstart+3*\a+\X,\ystart+3*\b+\Y);
        
        \draw[black] (\xstart,\ystart) rectangle (\xstart+\a,\ystart+\b);
        \draw[black] (\xstart+\a,\ystart) rectangle (\xstart+2*\a,\ystart+\b);
        \draw[black] (\xstart+2*\a,\ystart) rectangle (\xstart+3*\a,\ystart+\b);
        \draw[black,very thick] (\xstart,\ystart+\b) rectangle (\xstart+\a,\ystart+2*\b);
        \draw[black,very thick] (\xstart+\a,\ystart+\b) rectangle (\xstart+2*\a,\ystart+2*\b);
        \draw[black,very thick] (\xstart+2*\a,\ystart+\b) rectangle (\xstart+3*\a,\ystart+2*\b);
        \draw[black] (\xstart,\ystart+2*\b) rectangle (\xstart+\a,\ystart+3*\b);
        \draw[black] (\xstart+\a,\ystart+2*\b) rectangle (\xstart+2*\a,\ystart+3*\b);
        \draw[black] (\xstart+2*\a,\ystart+2*\b) rectangle (\xstart+3*\a,\ystart+3*\b);
        \filldraw[thick, top color=white,bottom color=gray!\bg!] (\xstart,\ystart+\b) rectangle (\xstart+\a,\ystart+2*\b);
        \filldraw[thick, top color=white,bottom color=gray!\bg!] (\xstart+\a,\ystart+\b) rectangle (\xstart+2*\a,\ystart+2*\b);
        \filldraw[thick, top color=white,bottom color=gray!\bg!] (\xstart+2*\a,\ystart+\b) rectangle (\xstart+3*\a,\ystart+2*\b);
        
        \draw[black,very thick] (\xstart-\X,\ystart-\Y) rectangle (\xstart+\a-\X,\ystart+\b-\Y);
        \draw[black,very thick] (\xstart+\a-\X,\ystart-\Y) rectangle (\xstart+2*\a-\X,\ystart+\b-\Y);
        \draw[black,very thick] (\xstart+2*\a-\X,\ystart-\Y) rectangle (\xstart+3*\a-\X,\ystart+\b-\Y);
        \draw[black] (\xstart-\X,\ystart+\b-\Y) rectangle (\xstart+\a-\X,\ystart+2*\b-\Y);
        \draw[black] (\xstart+\a-\X,\ystart+\b-\Y) rectangle (\xstart+2*\a-\X,\ystart+2*\b-\Y);
        \draw[black] (\xstart+2*\a-\X,\ystart+\b-\Y) rectangle (\xstart+3*\a-\X,\ystart+2*\b-\Y);
        \draw[black] (\xstart-\X,\ystart+2*\b-\Y) rectangle (\xstart+\a-\X,\ystart+3*\b-\Y);
        \draw[black] (\xstart+\a-\X,\ystart+2*\b-\Y) rectangle (\xstart+2*\a-\X,\ystart+3*\b-\Y);
        \draw[black] (\xstart+2*\a-\X,\ystart+2*\b-\Y) rectangle (\xstart+3*\a-\X,\ystart+3*\b-\Y);
        \filldraw[thick, top color=white,bottom color=gray!\bg!] (\xstart-\X,\ystart-\Y) rectangle (\xstart+\a-\X,\ystart+\b-\Y);
        \filldraw[thick, top color=white,bottom color=gray!\bg!] (\xstart+\a-\X,\ystart-\Y) rectangle (\xstart+2*\a-\X,\ystart+\b-\Y);
        \filldraw[thick, top color=white,bottom color=gray!\bg!] (\xstart+2*\a-\X,\ystart-\Y) rectangle (\xstart+3*\a-\X,\ystart+\b-\Y);
        
    \end{scope}
    
    
    \node[text width=0, rotate=\rot] at (\xtexts-\Xt-\fx,\ytexts+\Yt-\fy) {$\mathsf{F}\{3,1\}$};
    \node[text width=0, rotate=\rot] at (\xtexts+\at-\Xt-\fx,\ytexts+\bt+\Yt-\fy) {$\mathsf{F}\{3,2\}$};
    \node[text width=0, rotate=\rot] at (\xtexts+2*\at-\Xt-\fx,\ytexts+2*\bt+\Yt-\fy) {$\mathsf{F}\{3,3\}$};
    \node[text width=0, rotate=\rot] at (\xtexts-\aat-\Xt-\fx,\ytexts+\bbt+\Yt-\fy) {$\mathsf{F}\{2,1\}$};
    \node[text width=0, rotate=\rot] at (\xtexts-\aat+\at-\Xt-\fx,\ytexts+\bbt+\bt+\Yt-\fy) {$\mathsf{F}\{2,2\}$};
    \node[text width=0, rotate=\rot] at (\xtexts-\aat+2*\at-\Xt-\fx,\ytexts+\bbt+2*\bt+\Yt-\fy) {$\mathsf{F}\{2,3\}$};
    \node[text width=0, rotate=\rot] at (\xtexts-2*\aat-\gx-\Xt-\xnew,\ytexts+2*\bbt-\gy+\Yt) {\textbf{$\mathsf{G}\{1,1\}$}};
    \node[text width=0, rotate=\rot] at (\xtexts-2*\aat+\at-\gx-\Xt-\xnew,\ytexts+2*\bbt+\bt-\gy+\Yt) {\textbf{$\mathsf{G}\{1,2\}$}};
    \node[text width=0, rotate=\rot] at (\xtexts-2*\aat+2*\at-\gx-\Xt-\xnew,\ytexts+2*\bbt+2*\bt-\gy+\Yt) {\textbf{$\mathsf{G}\{1,3\}$}};
    
    \node[text width=0, rotate=\rot] at (\xtexts-\fx,\ytexts-\fy) {$\mathsf{F}\{3,1\}$};
    \node[text width=0, rotate=\rot] at (\xtexts+\at-\fx,\ytexts+\bt-\fy) {$\mathsf{F}\{3,2\}$};
    \node[text width=0, rotate=\rot] at (\xtexts+2*\at-\fx,\ytexts+2*\bt-\fy) {$\mathsf{F}\{3,3\}$};
    \node[text width=0, rotate=\rot] at (\xtexts-\aat-\gx-\xnew,\ytexts+\bbt-\gy) {\textbf{$\mathsf{G}\{2,1\}$}};
    \node[text width=0, rotate=\rot] at (\xtexts-\aat+\at-\gx-\xnew,\ytexts+\bbt+\bt-\gy) {\textbf{$\mathsf{G}\{2,2\}$}};
    \node[text width=0, rotate=\rot] at (\xtexts-\aat+2*\at-\gx-\xnew,\ytexts+\bbt+2*\bt-\gy) {\textbf{$\mathsf{G}\{2,3\}$}};
    \node[text width=0, rotate=\rot] at (\xtexts-2*\aat-\fx,\ytexts+2*\bbt-\fy) {$\mathsf{F}\{1,1\}$};
    \node[text width=0, rotate=\rot] at (\xtexts-2*\aat+\at-\fx,\ytexts+2*\bbt+\bt-\fy) {$\mathsf{F}\{1,2\}$};
    \node[text width=0, rotate=\rot] at (\xtexts-2*\aat+2*\at-\fx,\ytexts+2*\bbt+2*\bt-\fy) {$\mathsf{F}\{1,3\}$};
    
    \node[text width=0, rotate=\rot] at (\xtexts-\gx+\Xt-\xnew-\xnew,\ytexts-\gy-\Yt) {\textbf{$\mathsf{G}\{3,1\}$}};
    \node[text width=0, rotate=\rot] at (\xtexts+\at-\gx+\Xt-\xnew-\xnew,\ytexts+\bt-\gy-\Yt) {\textbf{$\mathsf{G}\{3,2\}$}};
    \node[text width=0, rotate=\rot] at (\xtexts+2*\at-\gx+\Xt-\xnew-\xnew,\ytexts+2*\bt-\gy-\Yt) {\textbf{$\mathsf{G}\{3,3\}$}};
    \node[text width=0, rotate=\rot] at (\xtexts-\aat-\fx+\Xt,\ytexts+\bbt-\fy-\Yt) {$\mathsf{F}\{2,1\}$};
    \node[text width=0, rotate=\rot] at (\xtexts-\aat+\at-\fx+\Xt,\ytexts+\bbt+\bt-\fy-\Yt) {$\mathsf{F}\{2,2\}$};
    \node[text width=0, rotate=\rot] at (\xtexts-\aat+2*\at-\fx+\Xt,\ytexts+\bbt+2*\bt-\fy-\Yt) {$\mathsf{F}\{2,3\}$};
    \node[text width=0, rotate=\rot] at (\xtexts-2*\aat-\fx+\Xt,\ytexts+2*\bbt-\fy-\Yt) {$\mathsf{F}\{1,1\}$};
    \node[text width=0, rotate=\rot] at (\xtexts-2*\aat+\at-\fx+\Xt,\ytexts+2*\bbt+\bt-\fy-\Yt) {$\mathsf{F}\{1,2\}$};
    \node[text width=0, rotate=\rot] at (\xtexts-2*\aat+2*\at-\fx+\Xt,\ytexts+2*\bbt+2*\bt-\fy-\Yt) {$\mathsf{F}\{1,3\}$};
    
    \node[text width=3.5cm] at (\xlabels-\al+2,\ylabels+\bl) {
    the 1st layer 
    of vector indices \\
    $\{1, \hdots, \ntp$\}};
    \node[text width=4cm] at (\xlabels+2,\ylabels) {the 2nd layer 
    of vector indices \\ $\{\ntp+1, \hdots, 2\ntp$\}};
    \node[text width=4cm] at (\xlabels+\al+2,\ylabels-\bl) {the 3rd layer 
    of vector indices \\ $\{2\ntp+1, \hdots, 3\ntp$\}};
    
    \filldraw [black] (\xdota,\ydota) circle (\r);
    \filldraw [black] (\xdota+\dw,\ydota-\dh) circle (\r);
    \filldraw [black] (\xdota+2*\dw,\ydota-2*\dh) circle (\r);
    \filldraw [black] (\xdota+3*\dw,\ydota-3*\dh) circle (\r);
    \filldraw [black] (\xdota+4*\dw,\ydota-4*\dh) circle (\r);
    \filldraw [black] (\xdota+5*\dw,\ydota-5*\dh) circle (\r);
    \filldraw [black] (\xdota+6*\dw,\ydota-6*\dh) circle (\r);
    \filldraw [black] (\xdota+7*\dw,\ydota-7*\dh) circle (\r);
    \filldraw [black] (\xdota+8*\dw,\ydota-8*\dh) circle (\r);
    
    \filldraw [black] (\xdota-\Xc,\ydota+\Yc) circle (\r);
    \filldraw [black] (\xdota+\dw-\Xc,\ydota-\dh+\Yc) circle (\r);
    \filldraw [black] (\xdota+2*\dw-\Xc,\ydota-2*\dh+\Yc) circle (\r);
    \filldraw [black] (\xdota+3*\dw-\Xc,\ydota-3*\dh+\Yc) circle (\r);
    \filldraw [black] (\xdota+4*\dw-\Xc,\ydota-4*\dh+\Yc) circle (\r);
    \filldraw [black] (\xdota+5*\dw-\Xc,\ydota-5*\dh+\Yc) circle (\r);
    \filldraw [black] (\xdota+6*\dw-\Xc,\ydota-6*\dh+\Yc) circle (\r);
    \filldraw [black] (\xdota+7*\dw-\Xc,\ydota-7*\dh+\Yc) circle (\r);
    \filldraw [black] (\xdota+8*\dw-\Xc,\ydota-8*\dh+\Yc) circle (\r);
    
    \filldraw [black] (\xdotb,\ydotb) circle (\r);
    \filldraw [black] (\xdotb+\dw,\ydotb-\dh) circle (\r);
    \filldraw [black] (\xdotb+2*\dw,\ydotb-2*\dh) circle (\r);
    \filldraw [black] (\xdotb+3*\dw,\ydotb-3*\dh) circle (\r);
    \filldraw [black] (\xdotb+4*\dw,\ydotb-4*\dh) circle (\r);
    \filldraw [black] (\xdotb+5*\dw,\ydotb-5*\dh) circle (\r);
    \filldraw [black] (\xdotb+6*\dw,\ydotb-6*\dh) circle (\r);
    \filldraw [black] (\xdotb+7*\dw,\ydotb-7*\dh) circle (\r);
    \filldraw [black] (\xdotb+8*\dw,\ydotb-8*\dh) circle (\r);
    
    \filldraw [black] (\xdotb-\Xc,\ydotb+\Yc) circle (\r);
    \filldraw [black] (\xdotb+\dw-\Xc,\ydotb-\dh+\Yc) circle (\r);
    \filldraw [black] (\xdotb+2*\dw-\Xc,\ydotb-2*\dh+\Yc) circle (\r);
    \filldraw [black] (\xdotb+3*\dw-\Xc,\ydotb-3*\dh+\Yc) circle (\r);
    \filldraw [black] (\xdotb+4*\dw-\Xc,\ydotb-4*\dh+\Yc) circle (\r);
    \filldraw [black] (\xdotb+5*\dw-\Xc,\ydotb-5*\dh+\Yc) circle (\r);
    \filldraw [black] (\xdotb+6*\dw-\Xc,\ydotb-6*\dh+\Yc) circle (\r);
    \filldraw [black] (\xdotb+7*\dw-\Xc,\ydotb-7*\dh+\Yc) circle (\r);
    \filldraw [black] (\xdotb+8*\dw-\Xc,\ydotb-8*\dh+\Yc) circle (\r);
    
    
    
    \draw [decorate,decoration={brace,amplitude=10pt,mirror},rotate=0]
    (-6,11) -- (-5,1.6) node [black,midway]{\footnotesize};
    
    \node[text width=3.5cm] at (-5,6.25) {$3\ntp$};
\end{tikzpicture}
\caption{The \textbfn{GG} structure.} \label{GG}
\end{figure}

The extended procedure \textbf{energies} evaluates all $\varepsilon$-perturbed values of energies:
\begin{listing}
\begin{lstlisting}
function e = energies(v,eps,mesh,patches,params,indx)
% components of deformation
v_cell = createCellFromVector(v,mesh.dim);
% components of deformation on elements
v_elems = CellAtMatrixOfIndices(v_cell,mesh.elems2nodes);
% deformation gradients on elements
F_elems = evaluate_F(mesh,v_elems);

% deformations on patches
v_patches = CellAtMatrixOfIndices(v_cell,patches.elems2nodes);
% deformation gradients on patches
F_patches = CellAtMatrixOfIndices(F_elems,patches.elems);
        
v_patches_eps = cell(3,1);
e = zeros(3*numel(mesh.nodesMinim),size(eps,2));
for comp=1:size(eps,2)  % loop over epsilon perturbations
    % perturbations of deformations on patches
    v_patches_eps{1} = v_patches{1} + eps(comp)*patches.logical;
    v_patches_eps{2} = v_patches{2} + eps(comp)*patches.logical;
    v_patches_eps{3} = v_patches{3} + eps(comp)*patches.logical;
    
    % deformation gradients of perturbation
    GG = evaluate_GG(patches,v_patches_eps,F_patches);
    
    % densities
    densities_patches = densityGradientVector_3D(GG,params);  
    % energies
    e_patches = [patches.volumes; patches.volumes; patches.volumes].*densities_patches;
    % final energy values on patches
    csep = cumsum(e_patches);      
    e([1:3:end 2:3:end 3:3:end],comp) = [csep(indx(1)); diff(csep(indx))];
end
e = e(mesh.dofsMinim_local,:);   % dropping out the Dirichlet dofs
end
\end{lstlisting}
\end{listing}
{\color{review}The cells \textbfn{v\_elems} and \textbfn{F\_elems} are the same as in the first-gradient energy evaluation procedure from Subsection \ref{subs:fgenergy}.} The cell \textbfn{v\_patches\_eps} contains matrices
\begin{center}
\textbf{v\_patches\_eps$\{1\}$}, \textbf{v\_patches\_eps$\{2\}$}, \textbf{v\_patches\_eps$\{3\}$} \quad of size $\ntp \, \times \, 4$
\end{center}
that provide {\color{myart}the} restrictions of nodal deformations to all patches, but here these deformations are perturbed by the value of the central difference step $\varepsilon$. {\color{review}This cell is used for the construction of the key cell \textbfn{GG} of size $\dim \times \dim$ containing vectors
\begin{center}
\textbf{GG}$\{\j,\m\}$ \; of length $\dim \, \ntp$
\end{center}
storing the partial derivatives of deformations of the $\j$-th component with respect to the $\m$-th variable. The structure of \textbfn{GG} is in details displayed in Fig. \ref{GG}.}

\noindent
{\color{review}All three layers of vector indices have 
the same size of 
$\ntp$ entries and the following meaning:
\begin{itemize}
    \item[-] the 1st layer corresponds to the $\varepsilon$-perturbation of the component $\vv^{(1)}$,
    \item[-] the 2nd layer corresponds to the $\varepsilon$-perturbation of the component $\vv^{(2)}$,
    \item[-] the 3rd layer corresponds to the $\varepsilon$-perturbation of the component $\vv^{(3)}$.
\end{itemize}
}
\noindent
The cell \textbfn{G} has the same size as \textbfn{F}, but contains deformation gradient matrices corresponding to the deformations perturbed by $\varepsilon$.
The most important feature is using the values of the precomputed input cell \textbfn{F} for the efficient construction of the \textbfn{GG} cell. Note that if the numeric difference step $\varepsilon$ is added or subtracted from the first component of displacements, the deformation gradients of the rest two components remain the same and their values are already stored in the \textbfn{F} cell. \\
\noindent
A vector \textbfn{csep} of length $\d \ntp$ contains {\color{myart}the} cumulative sums of \textbfn{e\_patches} with elements
$$ \mbox{\textbf{csep}}(\n) = \sum\limits_{\g=1}^{\n} \mbox{\textbf{e\_patches}}(\g) \, . $$
{\color{myart}An} output matrix \textbf{e} of size $3 \nnf \times 2$ contains all energy contributions and for any $comp \in \{1,2\}$
\begin{equation}
        \mbox{\textbf{e}}(\n,comp) 
        = \sum\limits_{\g=p_{\n-1}+1}^{p_{\n}} \mbox{\textbf{e\_patches}}(\g)  
        = \textbf{csep}(p_{\n}) - \textbf{csep}(p_{\n}-1) \, .
\end{equation}
Note that vector \textbfn{e\_patches} changes its value inside the loop over {\color{myart}the} components $comp$. 
Therefore, the whole vector \textbf{e} is evaluated at once using the Matlab difference function \textbf{diff} with the input vector \textbfn{indx} of length $3\nnf$, where \textbf{indx}$(\n) = p_{\n} \, , \nfset$, with $p_{\n}$ defined in \eqref{pind1} and additionally {\color{myrev}in} \eqref{pind2}.

The procedure for the construction of \textbfn{GG} is listed below:

\begin{listing}
\begin{lstlisting}
function GG = evaluate_GG(patches,v_patches,F)
% deformation gradients on patches 
G = evaluate_F(patches,v_patches);

% deformation gradient on patches and on 3 components
GG = cell(3,3);
GG{1,1} = [G{1,1}; F{1,1}; F{1,1}];
GG{1,2} = [G{1,2}; F{1,2}; F{1,2}];
GG{1,3} = [G{1,3}; F{1,3}; F{1,3}];
GG{2,1} = [F{2,1}; G{2,1}; F{2,1}];
GG{2,2} = [F{2,2}; G{2,2}; F{2,2}];
GG{2,3} = [F{2,3}; G{2,3}; F{2,3}];
GG{3,1} = [F{3,1}; F{3,1}; G{3,1}];
GG{3,2} = [F{3,2}; F{3,2}; G{3,2}]; 
GG{3,3} = [F{3,3}; F{3,3}; G{3,3}];
end
\end{lstlisting}
\end{listing}

\subsection{Exact approach to evaluate $\nabla \Jgradv$}

Evaluation of the exact $\nabla \Jv$ requires {\color{myart}an} explicit deriving of every $\frac{\partial \Jv}{\partial v_{\n}} \, , \nfset$. {\color{myart}The} gradient of the linear part is trivial and is given by \eqref{linearTerm3}. Using \eqref{Denergy} one can write
\begin{equation}
    \frac{\partial \Jgradv}{\partial v_{\n}} = \sum_{\k=1}^{\nt} |T_{\k}| \, \frac{\partial}{\partial v_{\n}}W\big(\Fv|_{{T_{\k}}}\big) \, .
\end{equation} \label{dJdvn}
\noindent 
Note that the only elements whose energy contributions depend on $v_{\n}$ are those belonging to the $\ii$-th patch, where $\ii = \dnn$. Therefore, we can simplify the equation above as
\begin{equation}
    \frac{\partial \Jgradv}{\partial v_{\n}} = \sum_{\k=1}^{|\Tn|} |T_{\k}| \, \frac{\partial}{\partial v_{\n}}W\big(\Fv|_{T_{\k}}\big) \, .
\end{equation}
{\color{review}By using the chain rule one can write}
\begin{equation} \label{dWFv}
    \frac{\partial}{\partial v_{\n}}\WFv = \sum\limits_{\j=1}^{\d} \sum\limits_{\m=1}^{\dim} \frac{\partial W}{\partial \fjm}\big(\Fv\big) \dfjmvn (\vv) \, ,
\end{equation}
{\color{review}where assuming {\color{myart}the} Neo-Hookean density {\color{myrev}from} \eqref{neoHook}}
\begin{equation}
    \begin{split}
        \frac{\partial W(\F)}{\partial \fjm} = C_1\Big( \frac{\partial \I_1(\F)}{\partial \fjm} - \frac{2}{\det(\F)} \frac{\partial \det(\F)}{\partial \fjm} \Big) + 2 D_1\big(\det(\F)-1\big) \frac{\partial \det(\F)}{\partial \fjm} \, .
    \end{split}
\end{equation}
Using the row or {\color{myart}the} column expansion rule for calculating the determinant of a $3 \times 3$ matrix, one can express
\begin{equation}
    \frac{\partial \det(\F)}{\partial \fjm} = (-1)^{j+m} \det(\fjmsub) \, ,
\end{equation}
where $\fjmsub$ is {\color{myart}the} $2 \times 2$ submatrix of $\F$ given by dropping the $\j$-th row and the $\m$-th column of $\F$. Futhermore, $\frac{\partial \I_1}{\partial \fjm}$ can be also simplified by
\begin{equation}
    \frac{\partial \I_1(\F)}{\partial \fjm} = 2 \fjm \, .
\end{equation}
From Subsection \ref{subs:numeric} we already know how to assemble $\vv|_{T_{\k}}$ and $\Fv|_{T_{\k}}$. Here, in addition, we need to express $\frac{\partial \I_1(\F)}{\partial \F} \, , \frac{\partial \det(\F)}{\partial \F}$ and $\frac{\partial \F}{\partial v_{\n}}$.

Similarly to the numeric gradient evaluation, the exact gradient evaluation procedure uses the same precomputed structures \textbfn{v\_cell}, \textbfn{v\_elems} and \textbfn{F\_elems}.

\begin{listing}
\begin{lstlisting}
% components of deformation
v_cell = createCellFromVector(v,mesh.dim);
% components of deformation on elements
v_elems = CellAtMatrixOfIndices(v_cell,mesh.elems2nodes);
% deformation gradients on elements
F_elems = evaluate_F(mesh,v_elems);
\end{lstlisting}
\end{listing}

The main procedure for the exact gradient evaluation is provided below:

\begin{listing}
\begin{lstlisting}
% deformations on patches
v_patches = CellAtMatrixOfIndices(v_cell,patches.elems2nodes);
% deformation gradients on patches
F_patches = CellAtMatrixOfIndices(F_elems,patches.elems);

% deformation gradients of perturbation
GG = evaluate_GG(patches,v_patches,F_patches);

% densities
D_densities_patches = density_D_GradientVector_3D(GG,D_F,params);
% energies
D_e_patches = [patches.volumes; patches.volumes; patches.volumes].*D_densities_patches;
% final energy values on patches
csDep = cumsum(D_e_patches);
g = zeros(3*numel(nodesMinim),1);
g([1:3:end 2:3:end 3:3:end]) = [csDep(indx(1)); diff(csDep(indx))];
g = g(dofsMinim_local);
\end{lstlisting}
\end{listing}

\noindent
Here, \textbfn{D\_densities\_patches} is a vector of size $3\ntp \times 1$ containing all $\frac{\partial W(\F(\vv))}{\partial v_{\n}}$ for every $\nfset$ given by \eqref{dWFv}. Note that the function \textbf{density\_D\_GradientVector\_3D} uses the input $3 \times 3$ cell structure \textbfn{D\_F\_patches}, where \textbf{D\_F\_patches}$\{\j,\m\}(\n) = \frac{\partial \F_{\j,\m}}{\partial v_{\n}}$.

\begin{benchmark}\label{benchmark3}
The script \verb!benchmark3.m! evaluates approximations of $\nabla J_{grad}(\vv)$ on {\color{myart}the} sequence of {\color{myart}the} uniform mesh refinements defined in Benchmark \ref{benchmark2}. In order to provide {\color{review}higher accuracy of
evaluation times}, the energies are recomputed 10 times. Table \ref{tab3} provide evaluation times of {\color{myart}the} gradient approximations both in {\color{myart}the} exact and numerical case {\color{review}(the times for setting up \textbf{'mesh'} and \textbf{'patches'} structures are not included)}.

\newcolumntype{d}{>{\hsize=1\hsize}X}
\newcolumntype{Y}{>{\raggedleft\arraybackslash}X}
\newcolumntype{s}{>{\hsize=.5\hsize}Y}
\newcolumntype{z}{>{\hsize=1.4\hsize}Y}
\newcolumntype{Z}{>{\centering\arraybackslash}X }
\begin{table}[h]
    \centering
    \begin{tabularx}{0.8\textwidth}
    {s |Y |Y |Y }
   mesh level & number of free dofs & evaluation (10x) of exact $\nabla \Jv$: time  [s] & evaluation (10x) of num. $\nabla \Jv$: time  [s]  \\
     \hline 
 1 & 2133 &      0.12 &      0.08 \\ 
 2 & 11925 &      0.14 &      0.15 \\ 
 3 & 77517 &      1.18 &      1.51 \\ 
 4 & 554013 &     12.47 &     15.64 \\ 
 5 & 4178493 &    487.34 &    533.24 \\ 
    \end{tabularx}
    \vspace{0.3cm}
    \caption{Benchmark 3 - energy gradient evaluation times in 3D.}\label{tab3}
\end{table}
\end{benchmark}

\section{Application to practical energy minimizations}\label{sec:minimization}
For a practical energy minimization, the trust-region method \cite{conn2000} available in the MATLAB Optimization Toolbox is utilized. 
{\color{myrev} Standard stopping criteria (the first-order optimality, tolerance on the argument and tolerance on the function) equal to $10^{-6}$ are considered in all benchmarks.
}
The method also {\color{review}allows} to specify {\color{myart}the} sparsity pattern of the Hessian matrix 
$\nabla^2 J(v)$ in free degrees of freedom, i.e., only positions (indices) of nonzero entries. The sparsity pattern is directly given by a finite element discretization.
\begin{figure}[H]
\centering
\begin{minipage}[t]{0.99\textwidth}
\includegraphics[width=\textwidth]{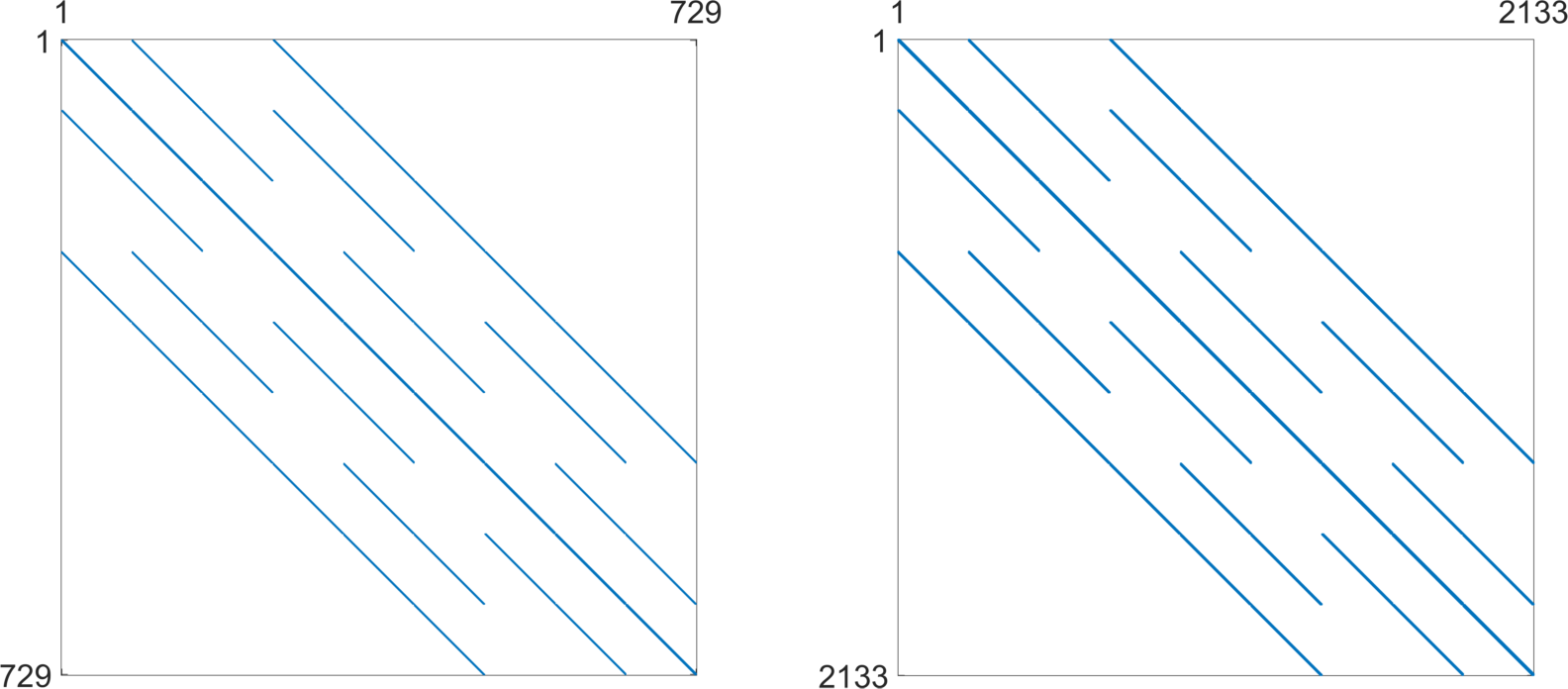}
\end{minipage}
\caption{The sparsity pattern of the 3D bar domain based on nodes connectivity (left) and its extension to the vector problem {\color{review}(right)}. }
\label{sparsity}
\end{figure}
\begin{example}
The sparsity pattern 
related to the 3D bar domain of Fig. \ref{mesh_torsion} is displayed in Fig. \ref{sparsity} (left). It corresponds to the FEM mesh not taking {\color{myart}the} Dirichlet boundary conditions into account. Therefore, it is of size $729 \times 729$. In practical computations, it is first extended from a scalar to {\color{myart}a} vector problem and then restricted to {\color{myart}the} \textbfn{mesh.dofsMinim} indices (right). Since there are 18 Dirichlet nodes fixed in all three directions, the corresponding number of rows and columns of the right hessian sparsity pattern is $(729-18)*3 = 2133$.
\end{example}

\subsection{Benchmark 4: time-dependent hyperelasticity in 3D}
We consider the elastic bar specified in Benchmark \ref{benchmark2} with a time-dependent nonhomogeneous Dirichlet boundary conditions on the right wall {\color{review}($x = l_x$)}. The torsion of the right wall is described by the formula \eqref{displacement} for {\color{review}$x = l_x = 0.4$}, where the rotation angle is assumed to be a time dependent function $\alpha = \alpha(t) = \alpha_{max} t/T$ for {\color{myart}a} sequence of integer discrete times $t \in \{1, 2, \dots, T\}$. Here we assume the final time $T = 24$ and the maximal angle of rotation $\alpha_{max} = 8\pi$ ensuring four full rotations of the right Dirichlet wall around the x-axis. Altogether we solve a sequence of $T$ minimization problems \eqref{minimization_general} with the Neo-Hook energy density \eqref{neoHook} and no loading ($\f = 0$).

\begin{figure}	
	\centering
	\begin{subfigure}{\textwidth}
		\centering
		\includegraphics[width=\textwidth]{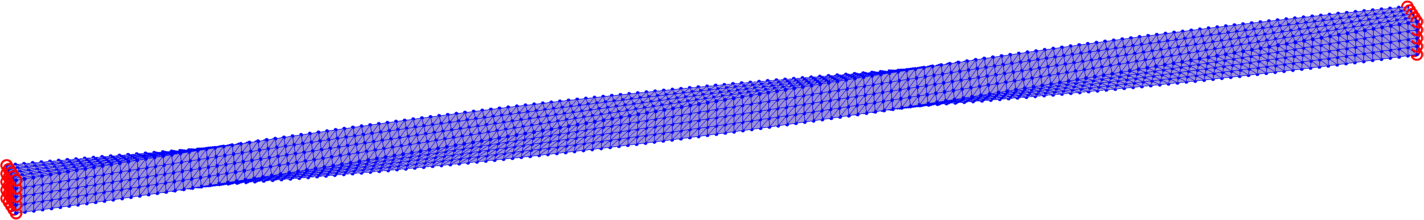}
		\vspace{-8mm}
		\caption{$t=3.$}	
	\end{subfigure}
	\vspace{-1mm}
	\begin{subfigure}{\textwidth}
		\centering
		\includegraphics[width=\textwidth]{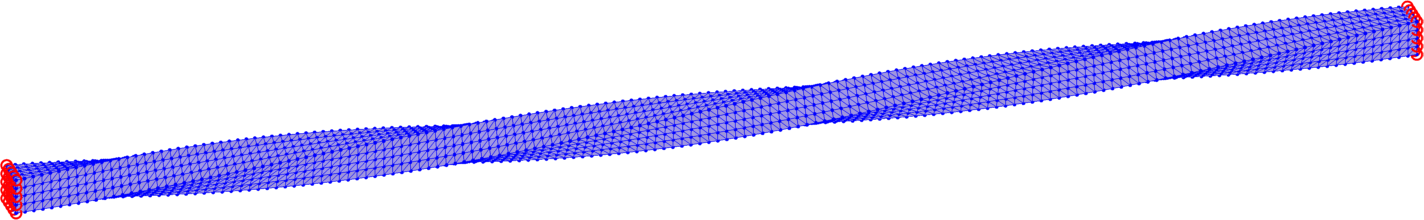}
		\vspace{-8mm}
		\caption{$t=6$.}	
	\end{subfigure}
	\vspace{-1mm}
	\begin{subfigure}{\textwidth}
		\centering
		\includegraphics[width=\textwidth]{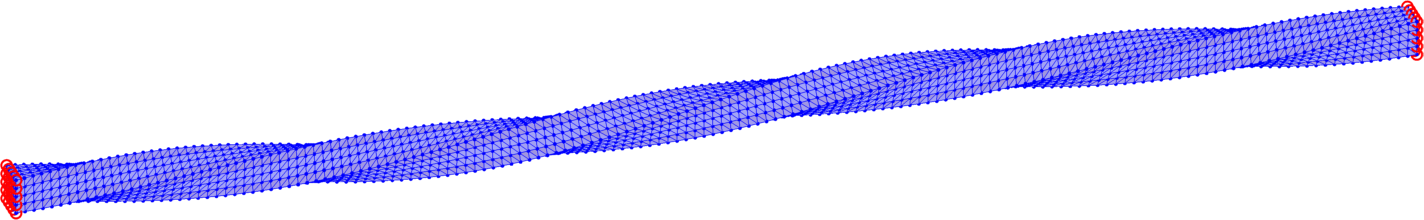}
		\vspace{-8mm}
		\caption{$t=9.$}	
	\end{subfigure}
	\vspace{-1mm}
	\begin{subfigure}{\textwidth}
		\centering
		\includegraphics[width=\textwidth]{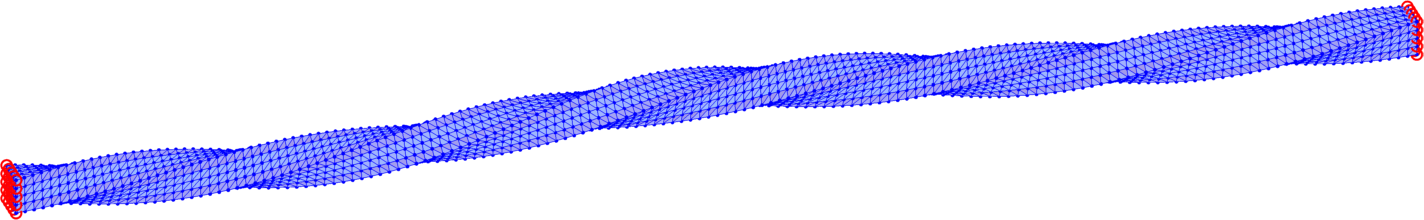}
		\vspace{-8mm}
		\caption{$t=12.$}	
	\end{subfigure}
	\vspace{-1mm}
	\begin{subfigure}{\textwidth}
		\centering
		\includegraphics[width=\textwidth]{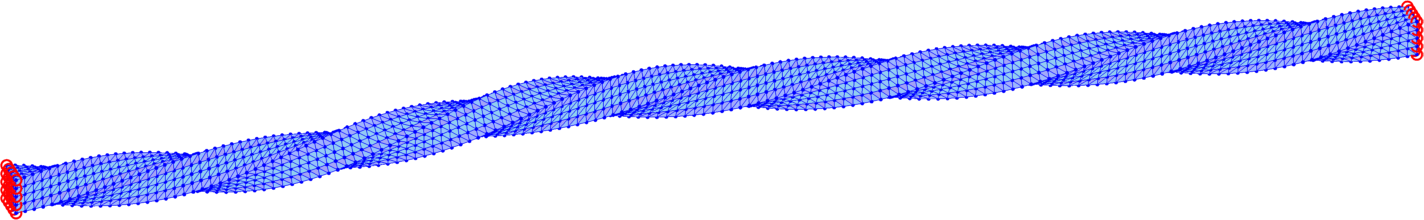}
		\vspace{-8mm}
		\caption{$t=15.$}	
	\end{subfigure}
	\vspace{-1mm}
	\begin{subfigure}{\textwidth}
		\centering
		\includegraphics[width=\textwidth]{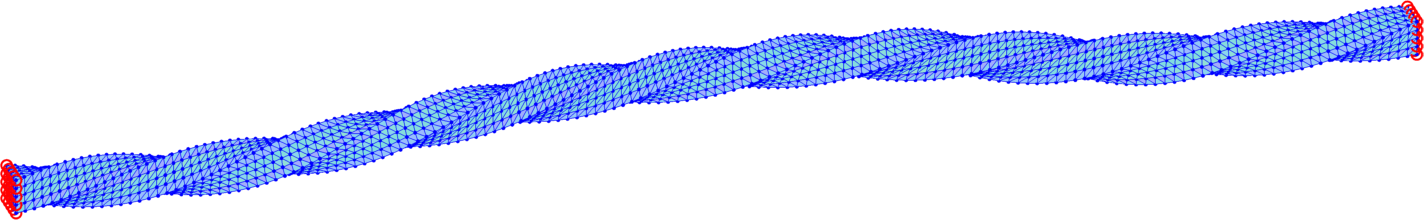}
		\vspace{-8mm}
		\caption{$t=18s.$}	
	\end{subfigure}
	\begin{subfigure}{\textwidth}
		\centering
		\includegraphics[width=\textwidth]{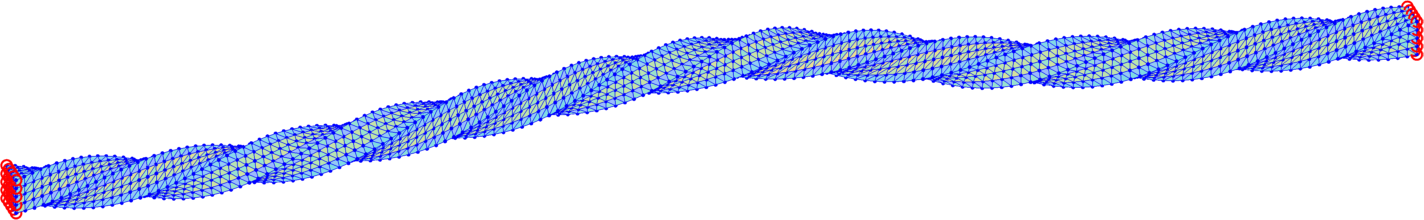}
		\vspace{-8mm}
		\caption{$t=21.$}	
	\end{subfigure}
	\begin{subfigure}{\textwidth}
		\centering
		\includegraphics[width=\textwidth]{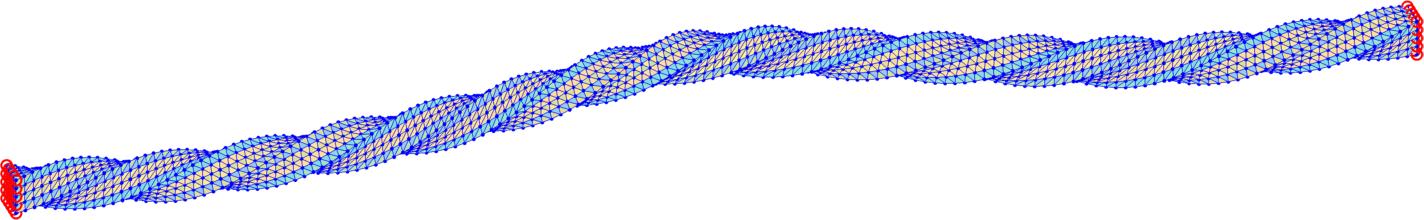}
		\vspace{-8mm}
		\caption{$t=24.$}	
	\end{subfigure}
	\caption{Benchmark 4 - deformation of the tetrahedal mesh subjected to the right time-rotating Dirichlet plane with the underlying Neo-Hookean density.}\label{fig:animation}
\end{figure}

The trust-region method accepts an {\color{review}initial deformation approximation}. For the first discrete time we run iterations from the identity $v(x)=x, x \in \Omega$ and for the next discrete time the minimizer of the previous discrete time problem serves as its initial approximation. We output the found energy minimizer for each minimization problem with some minimizers shown in form of {\color{myart}the} deformed meshes in Fig. \ref{fig:animation}. Additionally, we also provide the minimization time, the number of the trust-region iterations and the value of the corresponding minimal energy. 

The overall performance is explained in Table \ref{tab:ti3} for separate computations {\color{myrev}on 
tetrahedral meshes of levels 1, 2, 3} applying the numerical  gradient evaluation of Subsection \ref{subs:numeric} only {\color{myrev} with the choice $\varepsilon = 10^{-6}$} . We notice that {\color{myart}the} minimizations of finer meshes require only slightly higher number of iterations, which is acceptable. Comparison of {\color{myart}the values $\Jgradu$ in each line of Table \ref{tab:ti3} indicates convergence in space of {\color{myart}the} energy minimizers}. 

\begin{table}[h]
{\color{review}
    \centering
    \begin{tabular}{r | r r r | r r r| r r r}
&  \multicolumn{3}{c|}{level 1, 2133 free dofs} 
 & \multicolumn{3}{c|}{level 2, 11925 free dofs} 
 & \multicolumn{3}{c}{level 3, 77517 free dofs}\\  
  \hline
step 
& time [s] & iters & $\Jgradu$ 
& time [s] & iters & $\Jgradu$ 
& time [s] & iters & $\Jgradu$ \\
\hline
t=3 &      6.65 & 40 &    3.1177&     66.23 & 47 &    1.8244&   2148.62 & 93 &    1.4631 \\ 
t=6 &     10.42 & 60 &   12.4423&    117.74 & 76 &    7.2962&   1404.68 & 68 &    5.8533 \\ 
t=9 &     10.33 & 60 &   27.8993&    125.76 & 85 &   16.4079&   1441.84 & 73 &   13.1735 \\ 
t=12 &     18.43 & 104 &   49.5506&    102.11 & 68 &   29.1664&   1277.04 & 65 &   23.4286 \\ 
t=15 &     15.51 & 81 &   77.3859&    127.17 & 87 &   45.5677&   1130.02 & 58 &   36.6236 \\ 
t=18 &     18.31 & 97 &  111.3369&    110.65 & 75 &   65.5470&   1540.90 & 76 &   52.7641 \\ 
t=21 &     15.74 & 82 &  151.4946&     74.47 & 51 &   89.1672&   5523.64 & 290 &   71.7883 \\ 
t=24 &     16.37 & 91 &  197.7577&     91.79 & 63 &  116.3261&   2660.44 & 131 &   93.7238 \\
    \end{tabular}
    \caption{Benchmark 4 - performance of hyperelasticity minimizations in 3D. }\label{tab:ti3}
    }
\end{table}

\vspace{-5mm}

\subsection{Benchmarks 5 and 6: hyperelasticity and p-Laplacian in 2D}
Although our exposition is mainly focused on implementation details in 3D, a reduction to 2D ($\dim=2$) is straightforward. 

\begin{figure}[H]
\centering
\begin{minipage}[c]{0.4\textwidth}
\includegraphics[width=\textwidth]{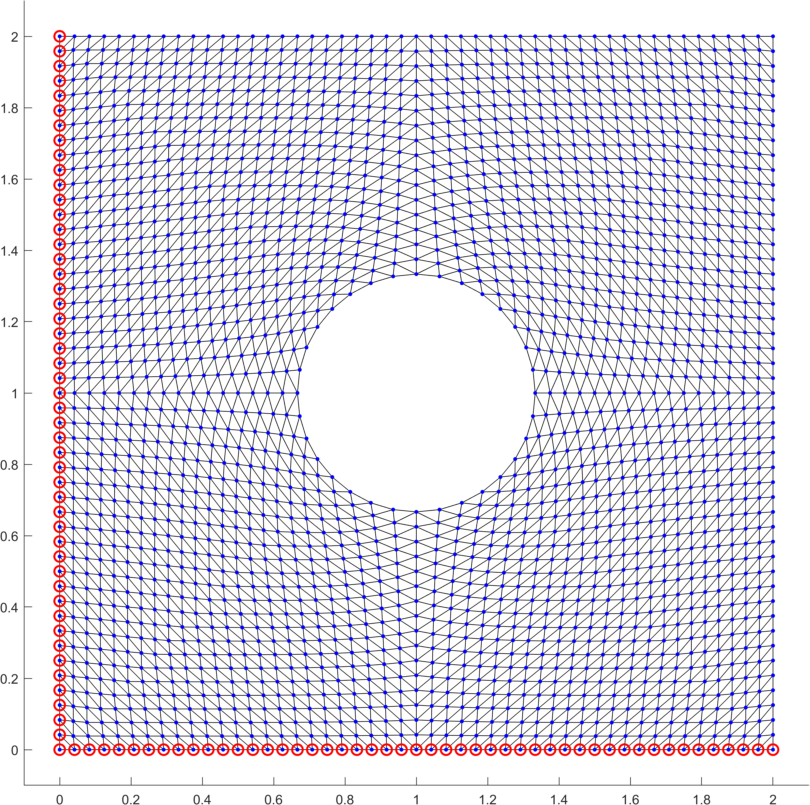}
\end{minipage}
\hspace{0.1\textwidth}
\begin{minipage}[c]{0.42\textwidth}
\includegraphics[width=\textwidth]{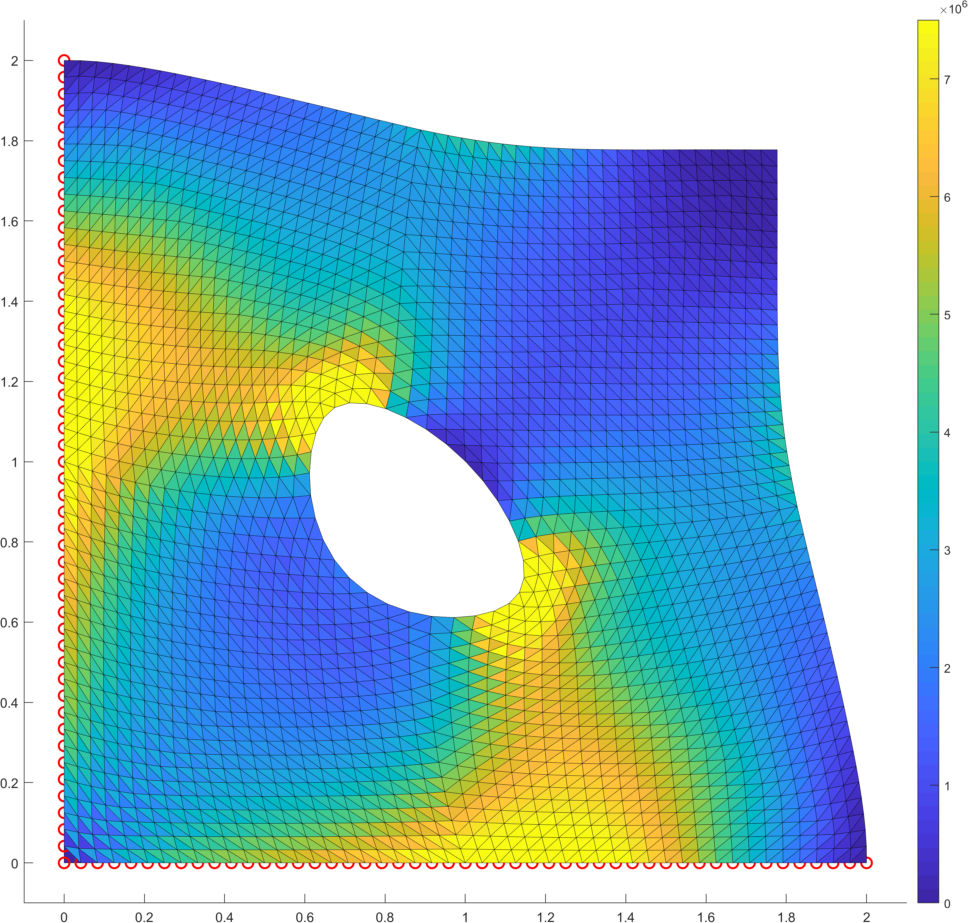}
\end{minipage} \:\:\:
\caption{Benchmark 5 - a triangulation of the square domain perforated by a hole (left) and its deformation with the underlying Neo-Hookean density (right).}
\label{elasticity_2D}
\vspace{8mm}
\centering
\begin{minipage}[c]{0.4\textwidth}
\includegraphics[width=\textwidth]{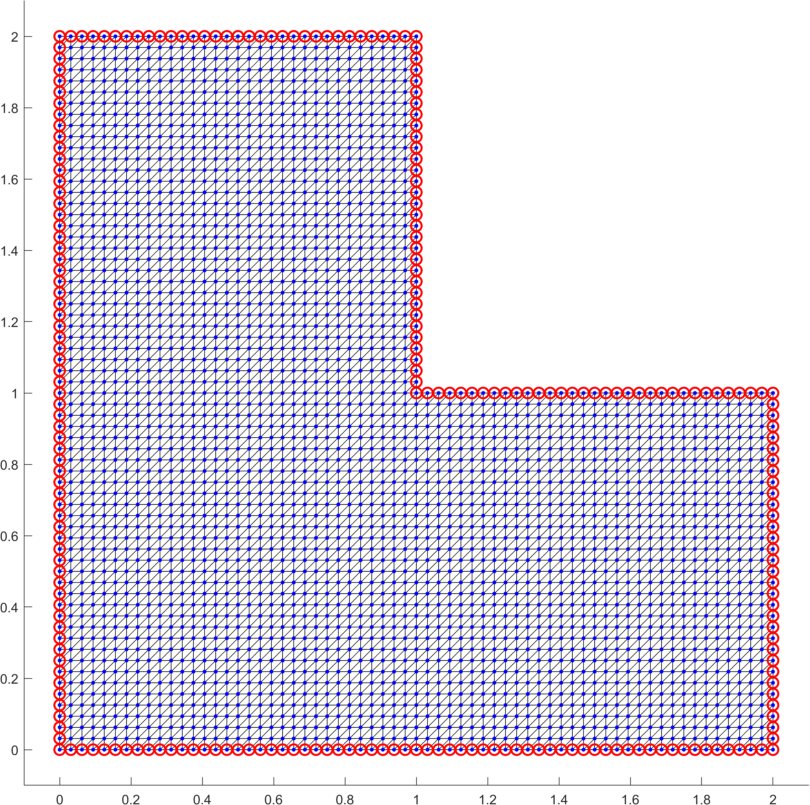}
\end{minipage}
\hspace{0.1\textwidth}
\vspace{0.05\textwidth}
\begin{minipage}[c]{0.42\textwidth}
\includegraphics[width=\textwidth]{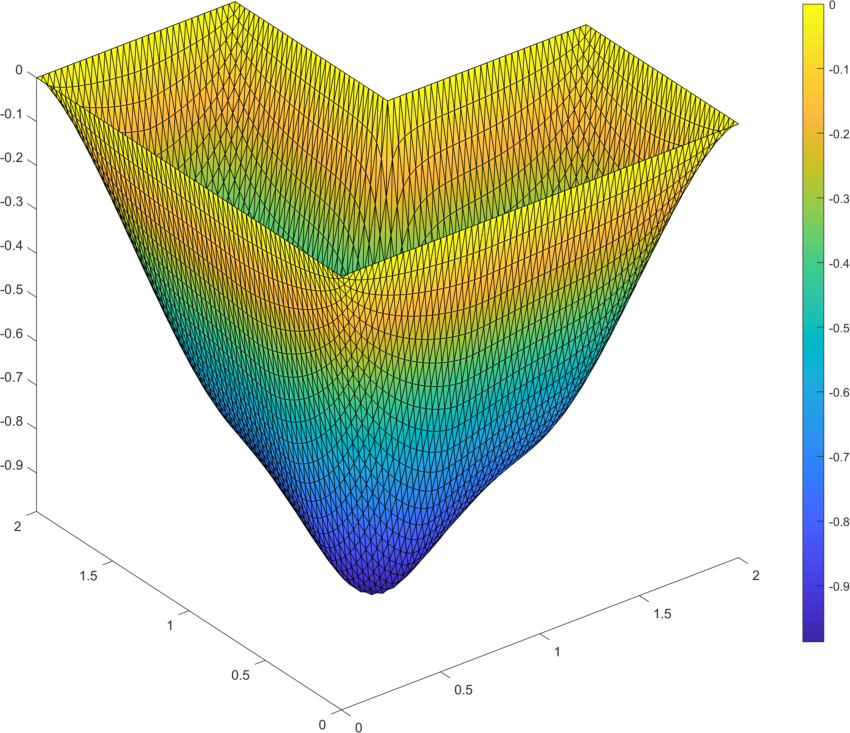}
\end{minipage}
\vspace{-0.5cm}
\caption{Benchmark 6 - a triangulation of the L-shape domain (left) and the solution p-Laplacian for the power $p=3$ and a constant loading $f(\x)=-10$ (right).}
\label{pLaplace_2D}
\end{figure}

As the first example we consider a square domain with lengths {\color{review}$l_x=l_y=2$} perforated by a hole with radius $r = 1/3$ (Fig. \ref{elasticity_2D} left) and whose center is located at the center of the square, subjected to zero Dirichlet boundary condition on the bottom and left edges, a constant loading {\color{myrev}$\f(\x) = (-3.5 \cdot 10^7, -3.5 \cdot 10^7)$} and the elastic parameters specified in Benchmark 2. 
{\color{myrev} We assume the choice $\varepsilon = 10^{-6}$} in the evaluation of the numerical gradient. The resulting deformation and the deformation gradient densities are depicted in Fig. \ref{elasticity_2D} (right).

As the second example we minimize the p-Laplacian energy functional \eqref{pLaplacian} over the L-shape domain (Fig. \ref{pLaplace_2D} left). The constant loading $f=-10$ is assumed together with zero Dirichlet boundary conditions on the full domain boundary. {\color{myrev}For the numerical gradient we assume $\varepsilon = 10^{-5}$}. The solution is displayed in Fig. \ref{pLaplace_2D} (right). Table \ref{tab:pLaplace} depicts performance of all options for the power $p=3$ and confirms faster evaluation times in comparison to our former contribution \cite{MMV}. This is mainly due to {\color{myart}the} improved  vectorization concepts here, due to the faster CPU and finally due to updates of the trust-region in-built implementation in the latest version of Matlab.

\begin{table}[h]
 \vspace{-0.5cm}
    \centering
    \begin{tabularx}{0.99\textwidth}
    {>{\raggedleft\arraybackslash}X
    >{\raggedleft\arraybackslash}X
    |>{\raggedleft\arraybackslash}X 
    >{\raggedleft\arraybackslash}X 
    >{\raggedleft\arraybackslash}X
    |>{\raggedleft\arraybackslash}X
    >{\raggedleft\arraybackslash}X
    >{\raggedleft\arraybackslash}X
    }
      & &  \multicolumn{3}{c|}{exact gradient} & \multicolumn{3}{c}{numerical gradient}  \\
      \hline
     level &  free dofs  &  time [s] & iters & $\Ju$ & time [s] & iters & $\Ju$ \\
\hline
 1 & 278 &      0.36 & 4 &   24.5745 &      0.13 & 4 &   24.5745 \\ 
 2 & 1102 &      0.15 & 4 &   24.3558 &      0.16 & 4 &   24.3558 \\ 
 3 & 4382 &      0.54 & 4 &   24.2957 &      0.54 & 4 &   24.2957 \\ 
 4 & 17470 &      2.35 & 5 &   24.2799 &      2.65 & 5 &   24.2799 \\ 
 5 & 69758 &     11.27 & 5 &   24.2758 &     12.39 & 5 &   24.2758 \\ 
 6 & 278782 &     83.27 & 6 &   24.2748 &     88.74 & 6 &   24.2748 \\
    \end{tabularx}
    \caption{Benchmark 5 - performance of hyperelasticity minimizations in 2D. True values of the energies $\Ju$ are multiplied by the scale factor $10^7$.}\label{tab:hyperelasticity}
    \vspace{0.5cm}
\end{table}
\vspace{-1mm}
\begin{table}[h]
 \vspace{-0.5cm}
    \centering
    \begin{tabularx}{0.99\textwidth}
    {>{\raggedleft\arraybackslash}X
    >{\raggedleft\arraybackslash}X
    |>{\raggedleft\arraybackslash}X 
    >{\raggedleft\arraybackslash}X 
    >{\raggedleft\arraybackslash}X
    |>{\raggedleft\arraybackslash}X
    >{\raggedleft\arraybackslash}X
    >{\raggedleft\arraybackslash}X
    }
      & &  \multicolumn{3}{c|}{exact gradient} & \multicolumn{3}{c}{numerical gradient}  \\
      \hline
     level &  free dofs  &  time [s] & iters & $\Ju$ & time [s] & iters & $\Ju$ \\
 \hline
 1 & 33 &      0.02 & 8 &   -7.5353 &      0.03 & 8 &   -7.5353 \\ 
 2 & 161 &      0.05 & 11 &   -7.9729 &      0.12 & 15 &   -7.9729 \\ 
 3 & 705 &      0.11 & 11 &   -8.1039 &      0.19 & 11 &   -8.1039 \\ 
 4 & 2945 &      0.30 & 11 &   -8.1445 &      0.56 & 12 &   -8.1445 \\ 
 5 & 12033 &      1.50 & 11 &   -8.1578 &      2.08 & 12 &   -8.1578 \\ 
 6 & 48641 &      6.30 & 12 &   -8.1625 &      9.48 & 12 &   -8.1625 \\ 
 7 & 195585 &     48.61 & 13 &   -8.1642 &     60.62 & 12 &   -8.1642 \\ 
 8 & 784385 &    617.92 & 22 &   -8.1649 &    672.04 & 16 &   -8.1649 \\ 
    \end{tabularx}
    \caption{Benchmark 6 - performance of p-Laplacian minimizations for $p=3$ in 2D.
    }\label{tab:pLaplace}
    \vspace{-0.5cm}
\end{table}

\subsection{{\color{myrev}Implementation remarks and outlooks}}
Assembly times in all benchmarks were obtained on a MacBook Air (M1 processor, 2020) with 16 GB memory running MATLAB R2021a. Our implementation is available 
at
\begin{center}
\url{https://www.mathworks.com/matlabcentral/fileexchange/97889} 
\end{center}
for download and testing. It is based on own codes of \cite{AnjamValdman2015,CSV2019,RahmanValdman2013} used primarily for assemblies of finite element matrices. It also utilizes the function \textbf{mcolon} from the {\color{review}reservoir} simulator \cite{KAL}. The names of mesh attributes were initially motivated by codes of \cite{ACF99} and further modified.

The 3D cuboid mesh is generated by the code of \cite{CSV2019} and 2D meshes (the square with the hole and the L-shape) by our own code.
Vectorized evaluations of gradients of basis functions are taken from \cite{AnjamValdman2015, RahmanValdman2013}. The general structure of the code was initially taken from \cite{MMV} and extended to the implementation of hyperelasticity in 2D and 3D. Many parts of this code (mainly related to the gradient evaluation) have been significantly improved since then.

{\color{review}The code is designed in a modular way allowing to add different scalar (e.g. Ginzburg-Landau \cite{ACF99}) and vector (e.g. topology optimization with elastoplasticity \cite{SAUS}, shape memory alloys \cite{Frost}) problems involving the first gradient energy terms.}

\vspace{-2mm}

\section*{Acknowledgement}
A. Moskovka was supported by the Strategy 21 of the CAS, program 23: City as a Laboratory of Change; Historical Heritage and Place for Safe and Quality Life and by the R$\&$D project 8J21AT001 Model Reduction and Optimal Control in Thermomechanics. J. Valdman  announces the support
of the Czech Science Foundation (GACR) through the grant GF19-29646L Large Strain Challenges in Materials Science.

\bibliographystyle{abbrv}

\end{document}